\documentclass[conference]{IEEEtran}
\IEEEoverridecommandlockouts
\usepackage{cite}
\usepackage{amsmath,amssymb,amsfonts}
\usepackage{algorithmic}
\usepackage{graphicx}
\usepackage{textcomp}
\usepackage{booktabs}
\usepackage{microtype}
\usepackage[font=small,skip=5pt]{caption}
\usepackage[table,xcdraw]{xcolor}

\usepackage{lmodern}
\usepackage[draft]{fixme}
\usepackage[utf8]{inputenc}
\usepackage{csvsimple}
\usepackage{amsmath}
\usepackage{float}
\usepackage{graphicx}
\usepackage{mathptmx}
\usepackage{url}
\usepackage{eqnarray}
\usepackage{graphicx}
\usepackage{caption}
\usepackage{subcaption}
\usepackage{stmaryrd}
\usepackage{balance}
\usepackage{caption}
\usepackage{color}
\usepackage{booktabs}
\usepackage{lmodern}
\usepackage{microtype}
\usepackage{hyperref}

\def\BibTeX{{\rm B\kern-.05em{\sc i\kern-.025em b}\kern-.08em
    T\kern-.1667em\lower.7ex\hbox{E}\kern-.125emX}}

\begin{document}

\title{Capstone: Mobility Modeling on Smartphones to Achieve Privacy by Design}

\author{\IEEEauthorblockN{Vaibhav Kulkarni, Arielle Moro, Bertil Chapuis, Beno\^{i}t Garbinato\\}
\IEEEauthorblockA{UNIL-HEC Lausanne\\
Distributed Object Programming Laboratory\\
firstname.lastname@unil.ch}
}

\maketitle

\begin{abstract}

Sharing location traces with context-aware service providers has privacy implications. 
Location-privacy preserving mechanisms, such as obfuscation, anonymization and cryptographic primitives, have been shown to have impractical utility/privacy tradeoff.    
Another solution for enhancing user privacy is to minimize data sharing by executing the tasks conventionally carried out at the service providers' end on the users' smartphones.
Although the data volume shared with the untrusted entities is significantly reduced, executing computationally demanding server-side tasks on resource-constrained smartphones is often impracticable.    
To this end, we propose a novel perspective on lowering the computational complexity by treating spatiotemporal trajectories as space-time signals. 
Lowering the data dimensionality facilitates offloading the computational tasks onto the digital-signal processors and the usage of the non-blocking signal-processing pipelines.  
While focusing on the task of user mobility modeling, we achieve the following results in comparison to the state of the art techniques: (i) mobility models with precision and recall greater than 80\%, (ii) reduction in computational complexity by a factor of 2.5, and (iii) reduction in power consumption by a factor of 0.5.
Furthermore, our technique does not rely on users' behavioral parameters that usually result in privacy-leakage and conclusive bias in the existing techniques.    
Using three real-world mobility datasets, we demonstrate that our technique addresses these weaknesses while formulating accurate user mobility models.\\

\end{abstract}


\section{Introduction}
\label{sec:intro}

A large amount of geolocation data is being ubiquitously collected, due to the advent of location-based services (LBS) and the pervasive nature of smartphones.
The personally identifiable information (PII) of users extracted from this data is crucial from the service providers perspective for offering personalized services. 
The accumulated data is used to constitute user specific, as well as collective mobility models, that encapsulate mobility behaviors.
Such models are used for a variety of applications such as location-based advertisements, traffic management and urban planning. 
However, when users share their location traces with third-party service providers, it exposes them to several privacy risks~\cite{Krumm2007InferenceAO}. 
Simple heuristics can be applied by curious adversaries to derive PII for blackmailing or stalking purposes~\cite{Gambs:2011:SMY:2019316.2019320}. 
Recent regulations such as the EU General Data Protection Regulation (GDPR){\footnote{GDPR: www.eugdpr.org}}, however, have placed stringent data acquisition and retention policies.
Article 25 ({\it{data protection by design and by default}}) lays out strict clauses for service providers, regarding the localization of computations and storage at the user's end whenever possible.{\footnote{Article 25 GDPR: gdpr-info.eu/art-25-gdpr}} 
A recent report claims that about 55\% of mobile applications do not currently comply with GDPR~\cite{SafeDk}.    
Therefore, user privacy consideration will be a key factor to determine the success and adoption of context-aware services in the coming years.

Several solutions have been proposed to address this issue in the context of LBS, including spatial cloaking~\cite{Gruteser2003AnonymousUO}, k-anonymity~\cite{Gedik2008ProtectingLP} and cryptographic primitives~\cite{Gahi2012PrivacyPS}.
Such techniques account for the optimization of the privacy/utility trade-off, where utility is often quantified in terms of the accuracy of the disclosed location traces~\cite{shokri2011quantifying}.  
However, such measures are still inefficient in deriving user mobility models with practically usable tradeoff~\cite{2011AnonymizationOL}.
Another category of solutions investigate data concealment, for example, Laplace perturbation, which encodes the trajectories with their Fourier transform coefficients~\cite{rastogi2010differentially}.    
However, data concealing and aggregation techniques are also exploitable due to the regularity and uniqueness of human mobility as shown by Xu et al.~\cite{xu2017trajectory}.

Orthogonal to the above solutions, a drastic privacy-preserving approach is to deploy the learning models directly on user's smartphones to train on their data without having to send it to the cloud.
An example of this approach is Google federated learning~\cite{google-federated-learning}. 
This model reverses the client/server relationship by enforcing the service providers to query for the required data from the model present on the user smartphone.
Finally, the query response can be processed in a trusted computing environment as illustrated in our previous work~\cite{Kulkarni2017PrivacyPreservingLS}.
Judicious scheduling in such systems ensures that learning occurs only when the device is completely idle~\cite{google-federated-learning}.
Thus, computational complexity and power consumption are the main concerns in making such a system practical.

In this paper, we adopt this type of approach consisting in restricting the mobility modeling task on the user's smartphone. 
Our approach in making such a system feasible is to treat spatiotemporal trajectories as signals.
To this end, we leverage the following key properties of spatiotemporal signals: (i) lower data magnitudes due to the reduced dimensionality, which is a direct application of Johnson–Lindenstrauss lemma (low-distortion embeddings of points from high-dimensional into low-dimensional Euclidean space), (ii) compressed representation, as the information is concentrated in a few spectral coefficients, and (iii) the ability to offload computationally intensive tasks to the digital-signal processors (DSPs) present in many smartphones.

\begin{figure}[t!]
\centering
\includegraphics[scale=0.63]{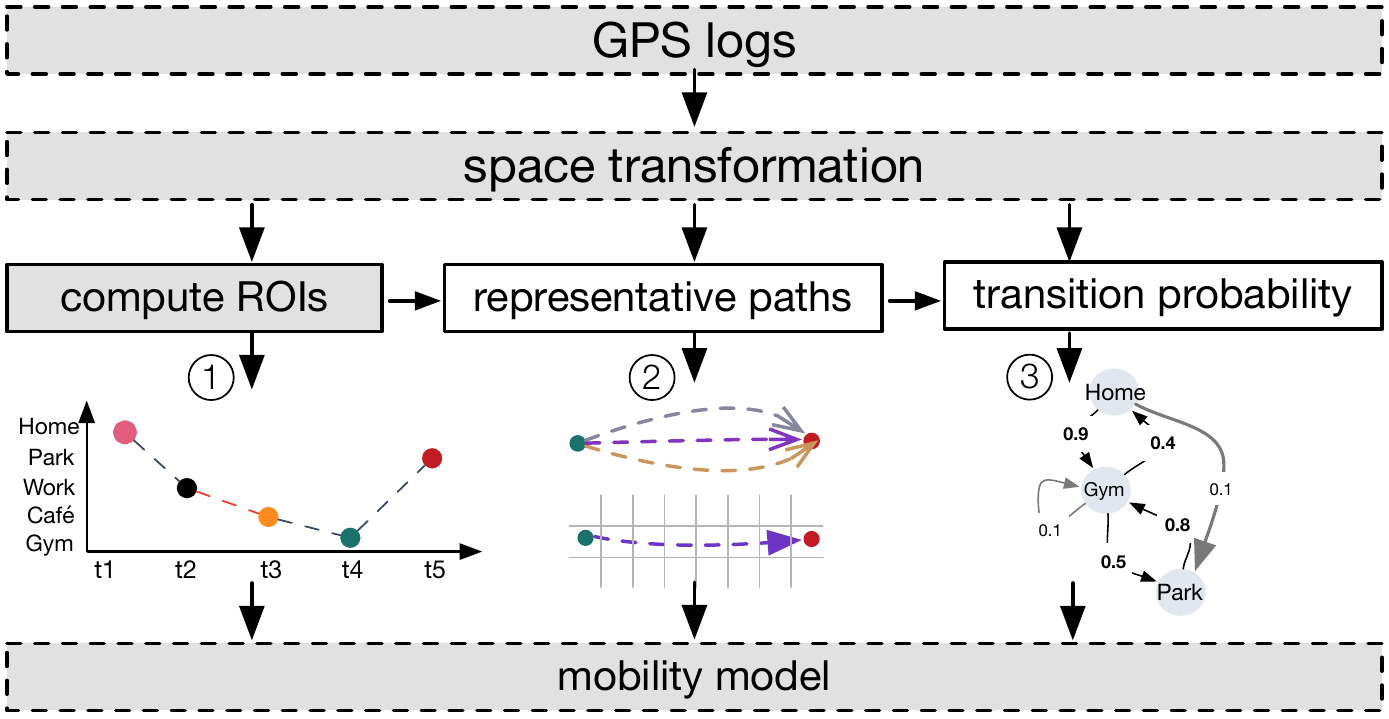}
\caption{Mobility modeling tasks: (1) computing ROIs, (2) estimating representative trajectories, and (3) computing transition probabilities.}
\label{fig:sys_mod}
\vspace{-15px}
\end{figure}

Traditionally, a mobility model is represented in terms of a directed graph, where the nodes correspond to the user's regions of interest (ROIs) and the edges correspond to the representative paths between the ROIs, weighted by the respective transition probabilities~\cite{Nguyen2011STEPSA}.    
Mobility modeling task is therefore composed of computing the ROIs, representative paths and the transition probabilities as depicted in Figure~\ref{fig:sys_mod}.
The current techniques used to perform the above tasks rely on an individual's behavioral parameters representing their mobility dynamics. 
However, these parameters act as side channels that can be used by malicious adversaries to infer an extended view of the whereabouts of a user appearing in an anonymous trajectory~\cite{Pellungrini2017ADM}.
Commonly used parameters such as minimum time period and maximum distance between two location coordinates can be used to de-anonymize aggregated spatiotemporal data~\cite{xu2017trajectory}.
We also show that reliance on these parameters result in conclusive bias and unfair comparisons of the efficacy of different techniques. 
We eliminate the dependence on the rigid parameter space and implement the proposed approach on a DSP chip to practically demonstrate the advantages.  
Our contributions in this context are as follows:

\begin{itemize}

	\item We present {\textbf{Capstone}}, a technique to construct a user mobility model using space-time signals. We divide our contributions in three distinct parts: (i) translating the noisy and non-uniformly sampled GPS trajectories into a continuous space-time signal, (ii) establishing a systematic relationship between the fundamental components of human mobility and the temporal-spectral units of the space-time signal, and (iii) a signal processing pipeline to extract user mobility model.

	\item We highlight the parameter curse present in the current techniques resulting in a strong conclusive bias and privacy leakage through experimental evaluation. We demonstrate the effectiveness of {\textbf{Capstone}} in addressing such drawbacks in addition to its suitability across a large variety of mobility datasets and disparate user mobility behaviors.

	\item Finally, by using three real-world mobility datasets, we show that \textbf{Capstone} achieves higher precision, lower complexity and reduction in power consumption as compared to the existing techniques, demonstrating its suitability to function on smartphones.  								 

\end{itemize}

We describe the privacy model and the problem statement in Section~\ref{sec:pri_mod} and Section~\ref{sec:problem_statement}. 
The three key contributions of {\textbf{Capstone}} are presented in Section~\ref{sec:mobsig},~\ref{sec:cross_domain} and~\ref{sec:system_design}.    
The drawbacks associated with the behavioral parameters are presented in Section~\ref{sec:a-priori} followed by the evaluation results and discussion in Section~\ref{sec:evaluation}.  
Finally, we present the related work in Section~\ref{sec:related_work} and conclude our paper in Section~\ref{sec:future_work}.

\section{Privacy and Attack Model}
\label{sec:pri_mod}

Our work focuses on privacy concerns associated with user mobility data, aggregated at the LBS provider's end.
We adopt the privacy by design principle, which demands inclusion of data protection convention from the onset of the system design.
More specifically, we base our model to comply with the EU data protection regulation.{\footnote{Art. 23: www.privacy-regulation.eu/en/article-23-restrictions-GDPR.htm}}

Location-based services are typically divided into two types: continuous and sporadic, depending on the exposure of user locations~\cite{shokri2011quantifying}.
In our case, we consider a continuous location exposure-based service, where the provider is assumed to be a passive adversary (i.e. honest-but-curious).    
We focus on converting such a continuous case, wherein the adversary can track users over time and space into a sporadic case, where a user explicitly grants location access to the adversary, only at discrete time instances.  
Thus, the adversary will know the geographical distribution of users over a considered region, but not their exhaustive movements.  
Therefore, by constructing the mobility model locally, our approach adopts a privacy by design principle, i.e., only sharing a summary/sketch of movements that is sufficient to use the service. 
We emphasize here that, we do not perform the space transformation (dimensionality reduction) as a means to encode the data in a format which directly preserves privacy. 
Instead, we simply leverage it to lower the computational complexity and power consumption.  
We also do not consider man-in-the-middle attacks or code injection attacks.

\section{Problem Statement}
\label{sec:problem_statement}

The key idea behind our work is: computation on spatiotemporal data using signal processing has inherent complexity and privacy benefits.
To this end, the central problem is construction of mobility models using the space-time signals in order to process and store user data locally.
Hereafter, we split this main problem statement into three sub-problems for clarity and set forth the requirements and challenges associated with each problem.\\ 

{\textbf{Problem 1: Mobility Signal Generation}}. Given a trajectory $T_u$ of an individual $u$, a temporally ordered sequence of tuples, such that, $T_u = \langle (l_1,t_1),(l_2,t_2)...(l_n,t_n)\rangle$, where $l_i = (lat_i,lon_i)$, the latitude-longitude coordinate pair and $t$, the timestamp such that $t_{i+1} > t_{i}$, translate $T_u$ into a 2-D signal $S_u(t)$, modeled as a function of changing distance with respect to time.
 
{\textbf{Requirements and Challenges}}. (i) Constructing a continuous graph from the noisy and non-uniformly sampled location trajectories, (ii) preserving all the key knowledge contained in the trajectory samples, and (iii) retaining the spatial locality between the discretized points.

\vspace{2px}

{\textbf{Problem 2: Signal Interpretation}}. Given a user's spatiotemporal signal $S_u(t)$, interpret and model the distinct signal elements in the temporal and spectral domain, i.e. local maxima/minima, rising/falling edges, static signal component, candidate frequencies, spectral coefficients and harmonics, with respect to human mobility behaviors.

{\textbf{Requirement and Challenge}}. In order to facilitate inter-domain switching, attach and validate a semantic meaning to each of the above signal components.

\vspace{2px}

{\textbf{Problem 3: Mobility Modeling}}. Given the signal $S_u(t)$ and the valid interpretation of each element, construct the user's mobility model in terms of a graph $G_u(ROI, Tr)$, where $ROI = \left\{ROI_1,ROI_2...ROI_n\right\}$ is the set of all the regions of interests belonging to $u$ and $Tr = \left\{(R_{12}, p_{12}),(R_{23},p_{23})...\right\}$ is the set of tuples where $R_{ij}$ and $p_{ij}$ denotes the representative path and the transition probability from $ROI_i$ to $ROI_j$.   
  
{\textbf{Requirement and Challenge}}. To extract all the distinct ROIs and the transitions without relying on any behavioral parameters to eliminate conclusive bias, privacy leakage and facilitate applicability across diverse datasets. 
\section{From Trajectories to Signals}
\label{sec:mobsig}

In this section we address {\textbf{Problem 1}}, i.e. translating the noisy GPS trajectories into a continuous signal, that can be processed. 

\subsection{Preprocessing}

The imperfections in the geolocation sensors and network failures often result in noisy and non-uniformly sampled trajectory points, thus hindering the process of generating a smooth and continuous signal. 
Therefore, to make the scheme robust, we first filter and de-noise the incoming location traces. 
Since the noise is not symmetrically distributed, applying averaging and median techniques does not solve this problem. 
Additionally, the noisy components reside at high frequencies and do not contain any sharp pulses, hence we apply a standard convolution-based low-pass filter.  
Next, we employ semivariance interpolation to obtain uniformly sampled location points. 
This interpolation scheme uses a moving average construction and conceals the incoming data about the spatial variance of the past trajectory points\cite{Kang2004ExtractingPF}.

\subsection{Space Discretization}

\begin{figure}[t!]
\centering
\includegraphics[scale=0.63]{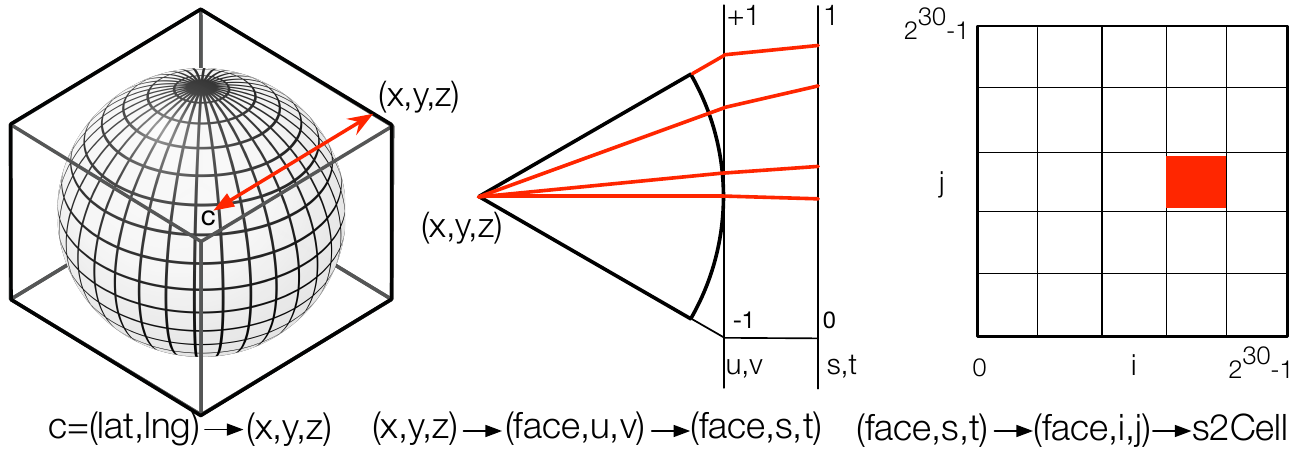}
\caption{Projecting a coordinate pair onto the grid with Google S2.}
\label{fig:dist_gs2}
\vspace{-15px}
\end{figure}

\begin{figure*}[!htbp]
    \centering
      \begin{subfigure}[b]{0.28\textwidth}
        \includegraphics[width=\textwidth]{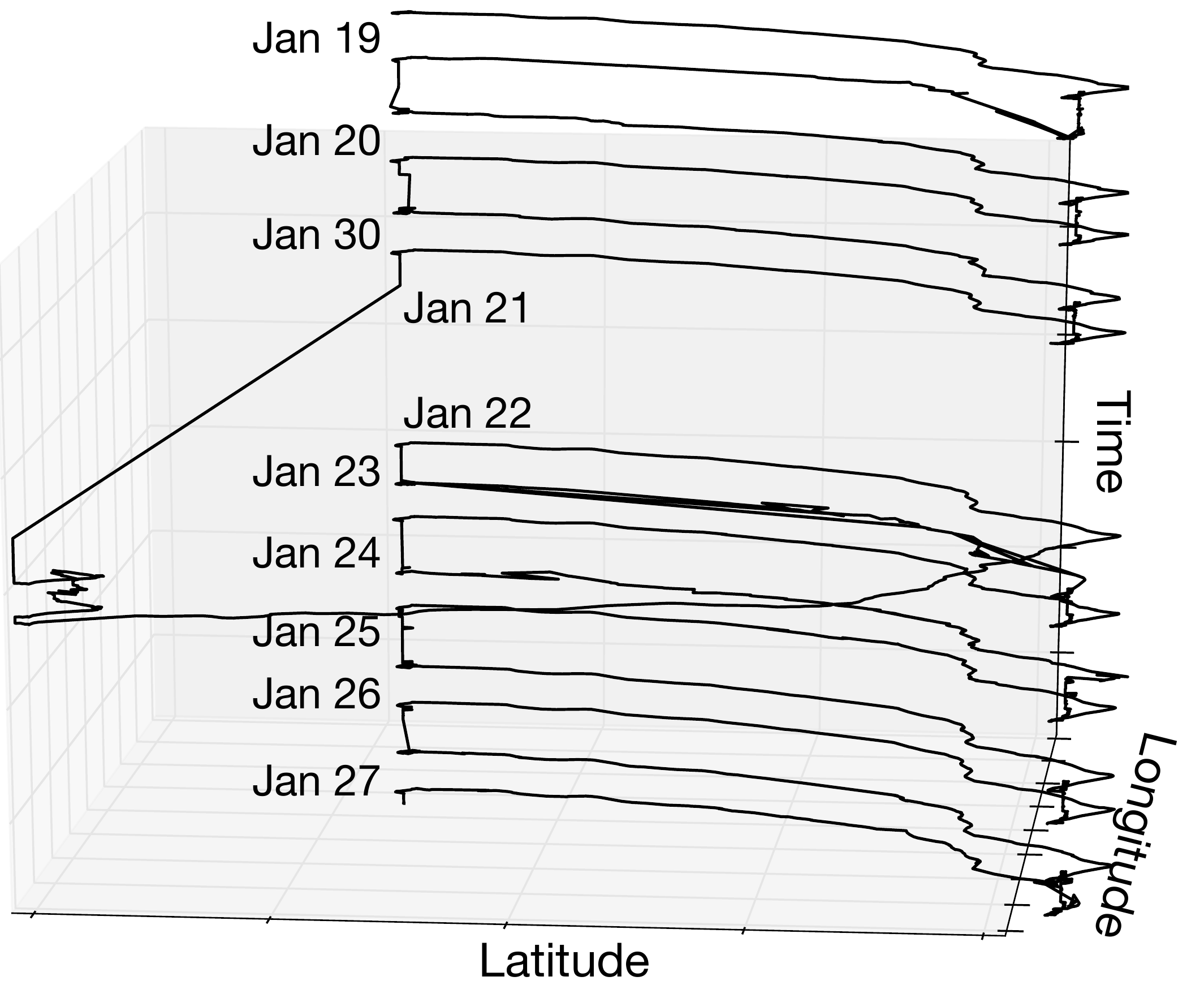}
        \caption{Visualization of a ten-day trajectory}
        \label{fig:a1}
    \end{subfigure}
    \begin{subfigure}[b]{0.33\textwidth}
        \includegraphics[width=\textwidth]{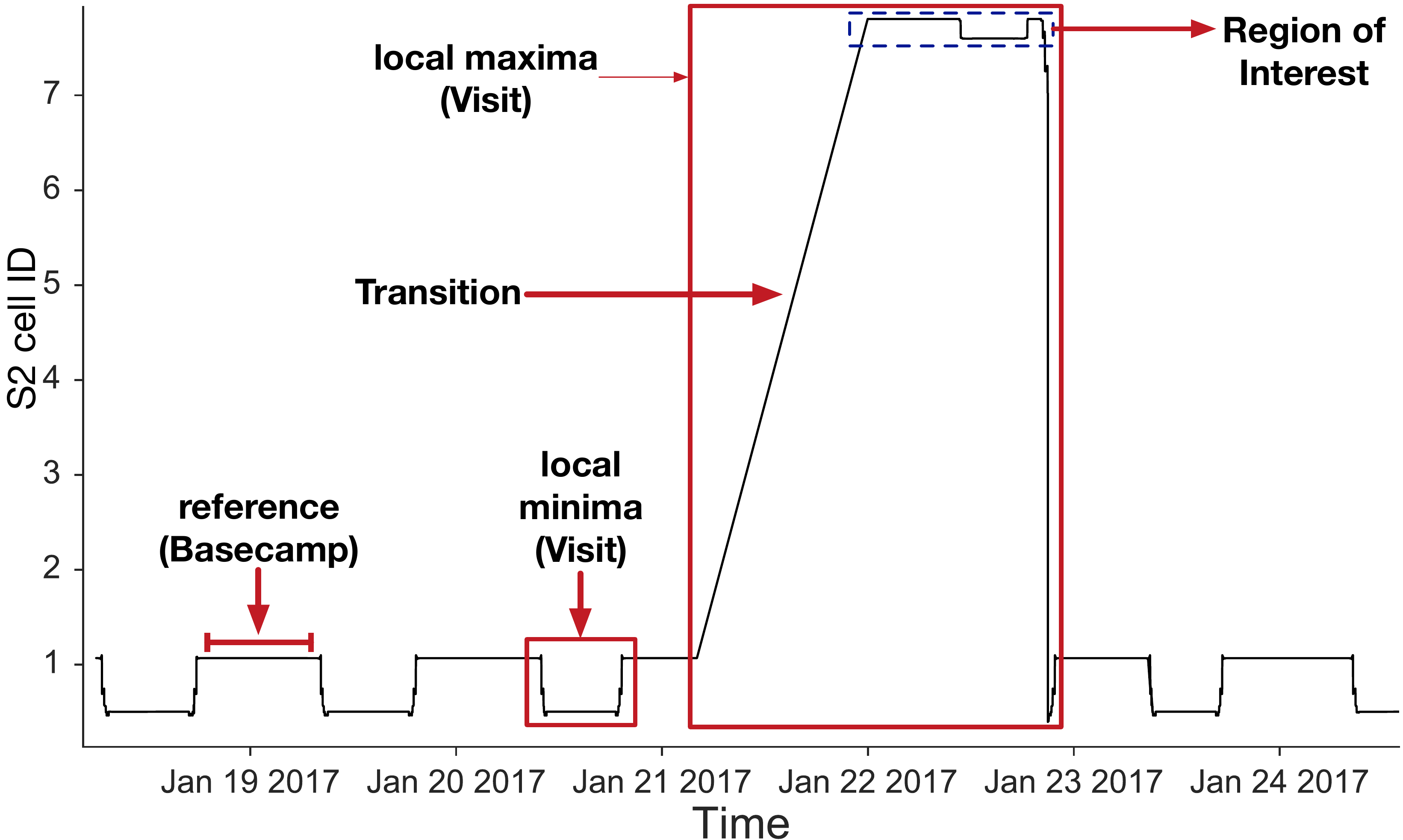}
        \caption{Visualization of the space-time signal}
        \label{fig:a2}
    \end{subfigure}
        \begin{subfigure}[b]{0.37\textwidth}
        \includegraphics[width=\textwidth]{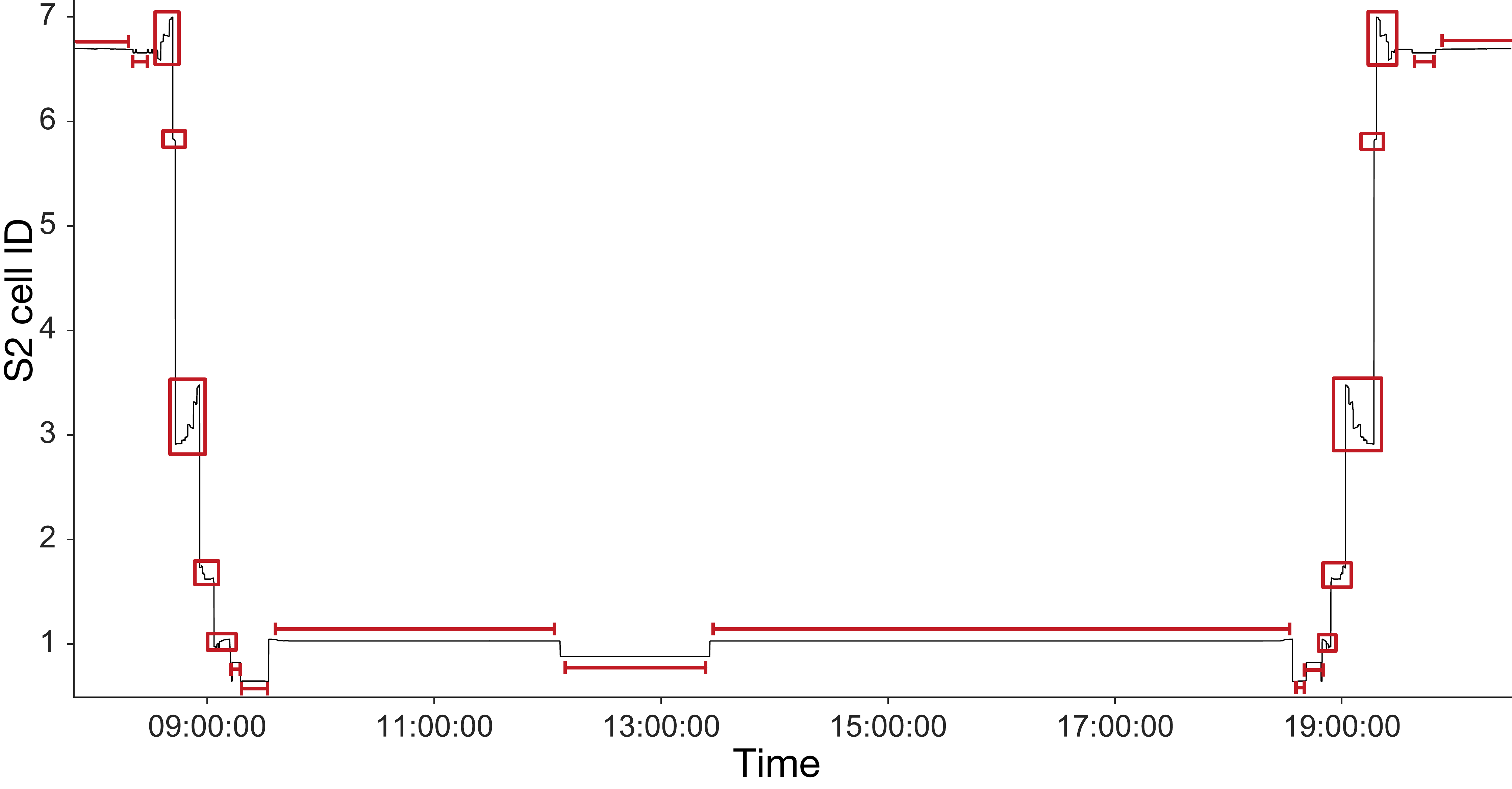}
        \caption{Visualizing one day's space-time signal}
        \label{fig:a3}
    \end{subfigure}
    \caption{Visualizing the user's movements as a trajectory and as space-time signal. The red rectangles and lines denote the ROIs.}
    \label{fig:a}
\vspace{-10px}    
\end{figure*}

The filtering and interpolation results in uniformly sampled de-noised trajectories. 
The process eliminates any bursty coordinates, however, some amount of white noise can still be present. 
The next step is space discretization, for which we rely on the Google S2 Library.{\footnote{Google S2: https://s2geometry.io/}}
It performs a hierarchical decomposition of the earth sphere into compact cells and superimposes a spatial region/point on to one of the cells.
Each cell is represented by exactly the same area and provides sufficient resolution for indexing the geographic features.    
In short, the library operates by first enclosing the earth in a cube.
It then projects the spatial region onto the face of the cube, builds a quad-tree on each face and selects the quad-tree cell that contains the projection of that region.
In the first step, the point $c=(lat,lon)$ in Figure~\ref{fig:dist_gs2} is transformed into $(x,y,z)$ after projecting it on the cube.
As the cells on the cube have different sizes when mapped back to the sphere, a non-linear transform is performed, i.e., $(u,v)$ is transformed to $(s,t)$ before discretizing the point by superimposing it on the grid and retrieving the respective {\it{Cell ID}}.
The cells are then enumerated on a Hilbert curve, preserving the spatial locality of the points~\cite{Moon2001AnalysisOT}.
The 64 bit {\it{Cell ID}} has 3 bits that encode the cube's face and the remaining 61 bits encode the position of this cell along he Hilbert curve.   
The resulting spatiotemporal signal can be denoted as $S(t) = \langle (c_1,t_1),(c_2,t_2)...(c_n,t_n)\rangle$, where $c_i$ is the {\it{Cell ID}} and $t_i$ the timestamp. 
The {\it{Cell IDs}} ensure that, each of the discrete point connects to the other to obtain a continuous graph, thus preserving the spatial locality between the individual points. 
The 3-D trajectories, shown in Figure~\ref{fig:a1} thus translate to 2-D space-time signals depicted in Figure~\ref{fig:a2}.

\section{Signal Interpretation}
\label{sec:cross_domain}

In this section, we address {\textbf{Problem 2}}, i.e., interpreting the distinct signal elements in time and frequency domain. 
The theoretical constructs proposed in this section are validated using three mobility datasets (discribed in Section~\ref{sec:evaluation}).  

\subsection{Temporal Domain}

A periodic signal $S(t)$ is typically represented as $S(t+n.T)$ for all time $t$ and the periodic component $T$, where $n.T$ is the period of the signal.
Although, human mobility is characterized by distinct regularity~\cite{Song2010LimitsOP}, the space-time signal $S(t)$ is not perfectly periodic.  
As the mobility patterns do not have the same mean periods, $S(t)$ can be considered as almost periodic~\cite{Trajkovic2009ModellingAF} and represented as $S(t) = S(t+n.T(t))$, where the fundamental period $T$, can change over time. 
It has two main components: (i) a static element and, (ii) rising/falling edges as seen in Figure~\ref{fig:a2}. 
In this work, the static element is treated as a {\it{reference}}, that corresponds to the user's {\it{basecamp}}. 
We refer to the {\it{basecamp}} as a place having maximum user time occupancy (typically the home or work place). 
It is represented in the time domain as $Mo(S(t))$, i.e., simply the {\it{Mode}} value of the signal, that correlates with the most frequent location in the user's trajectory.

This reference signal is accompanied with local maxima and minima. 
The user movements revolve around this reference with an element of deviation.
This is viewed as the presence of basic noise with a general mean.   
A user's ROI visit, thus corresponds to the local maxima/minima present in the signal and their amplitude correlates to the distance from the {\it{basecamp}} (or another ROI). 
A set of distinct ROIs can thus be obtained by selecting only the maxima/minima with distinct amplitudes. 
The maxima/minima significantly deviate from the noise and the reference element, and therefore are distinctly identifiable. 
A ROI visit can be expressed in terms of the local maxima at time $t_m$ as given in Equation~\ref{eq:roi_time}.
Here, $t_{m}-t_{e}$ and $t_{m}+t_{e}$ correspond to ROI visit outset and end times.

\vspace{-18px}

\begin{equation}
	Visit = \bigcup\limits_{t_e=-e}^{e} \lbrace S(t_m+t_e)\mid t_m, t_e \in t, S(t_m\pm t_e)>S(t) \rbrace
	\label{eq:roi_time} 
\end{equation}

Each maxima/minima can be decomposed into its constituents: (i) rising/falling edge, and (ii) local static element. 
It can be represented as $R_{edge} + LS_{static} + F_{edge}$, where $R_{edge}$ and $F_{edge}$, the rising and the falling edges correspond to the transition component and $LS_{static}$, the local static element maps to the user's ROI. 
A ROI visit can therefore be represented as $Visit = Tr + ROI$.
With respect to the spatial region, $Tr$ (transition) contains the knowledge of the trajectory traversed between the ROIs. 
Whereas, the $ROI$ holds the information regarding the spatial extent of the region of interest.

\subsection{Frequency Domain}

The spatiotemporal signal is inherently non-stationary, i.e., the frequency content of the signal changes over time.
Therefore, it needs to be processed as a short-term signal where it can be assumed as quasi-stationary.
Applying autocorrelation and power spectral density (PSD) analysis of such a signal extracts the candidate visitation periods.
As this signal can be viewed as a superposition of multiple periodic elements, each one corresponds to the visitation cycle associated with a distinct ROI. 
Applying a discrete cosine transform (DCT) further isolates the signal into fine grained constituents that correspond to the different movement patterns of a user. 
The periodicity associated with a single ROI visit corresponds to the frequency of one complex sinusoid and is represented as Equation~\ref{eq:single_freq}.  
\vspace{-5px}
\begin{equation}
	periodicity = \sum_{n=0}^{N} C_{i,n}.e^{ji\theta_n}
	\label{eq:single_freq}
\end{equation}  
     
In this equation, $C_{i,n}$ is the spectral coefficient that can change over time with respect to a time-dependent parameter $\theta_n$ bounded by $N$ and associated with a single frequency component $i$. 
Therefore, the complete set of periodicities associated with all the ROIs of a user can be derived by using Equation~\ref{eq:almost_periodic}.   
Here, the number of frequencies are restricted to an unknown but finite number $P$, the fundamental frequency $\theta_n$ and the spectral coefficients $C_p$, which can drift with time.

\begin{equation}
ROI_{periodicity} = \bigcup\limits_{p=0}^{P}\sum_{n=0}^{N}C_{p,n}.e^{jp\theta_n}
\label{eq:almost_periodic}
\end{equation}

An important property of DCT, which makes processing large magnitudes of trajectories viable on smartphones is its high degree of compaction. 
A DCT can provide a representation of the original signal by using a relatively small set of coefficients~\cite{Potluri2014Improved8A}.  
This is a highly desirable property when it comes to computing on resource constrained platforms, as it reduces the data storage requirements by storing only the coefficients that contain significant amounts of energy.
These coefficients retain the key signal information in a compressed state, such as the visitation periodicity and distance associated with the transitions.

In order to interpret the frequency domain components, we performed a modified discrete cosine transform (MDCT) over a fixed-size window (24 hours).
We correlate the scaled coefficients with the time domain signal and deduce that the low frequencies best describe the cycles and periods in the mobility traces. 
On the other hand, the high frequencies mostly contain noise, and can be eliminated. 
The abscissa represents the frequency of visitation (dominant periods), whereas the ordinate corresponds to the distance.
Furthermore, we also infer that the lower frequencies reside at higher distances and depict a periodic behavior and the harmonics represent the time-shifted versions of the ROI visits.

\section{Mobility Modeling}
\label{sec:system_design}

In this Section, we address {\textbf{Problem 3}}, i.e., mobility modeling and describe {\textbf{Capstone's}} system design and implementation. 
A mobility model consists of three main components: (i) ROIs, (ii) representative paths, and (iii) transition probabilities. 
Once all the ROIs are computed, the process of obtaining the representative paths and the probabilities is detailed in our previous work~\cite{Chapuis2016CapturingCB, Kulkarni:2016:MMP:3003421.3003424}.   
Constructing a representative path is essentially a procedure to efficiently extract the set of {\it{Cell IDs}} that best describes the trajectory between the two ROIs.
The transition probabilities can be computed by using mobility Markov chains~\cite{Kulkarni:2016:MMP:3003421.3003424}.

Following the discussion in Section~\ref{sec:cross_domain}, it is evident that the problem of constructing the mobility model, is essentially detecting the local maxima and minima (henceforth referred to as a 'peak') contained in the signal. 
This step is followed by isolating a peak into its constituents i.e., the set of cells associated with the ROI and the set of cells constituting the representative transition path. 
The latter is provided as an input to our technique~\cite{Chapuis2016CapturingCB} that extracts the path from this set. 
The system should be able to heuristically compute the following components upon which we base our design:
\begin{enumerate}
	\item peak start and end positions, to determine the ROI visit entry and end times;
	\item peak height, through which the distance travelled from the basecamp is calculated;
	\item peak width, to compute the total area and time spent at a given location;
	\item peak separation into travel time and stay time.
\end{enumerate}

\subsection{Visit Detection and Isolation}

\begin{figure}[t]
\centering
\includegraphics[scale=0.45]{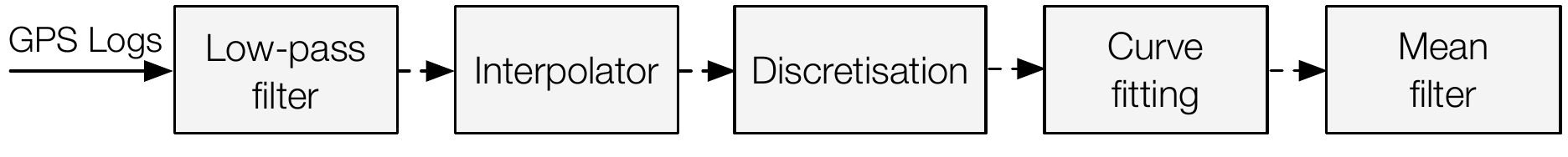}
\caption{Preprocessing steps.}
\label{fig:preprocess}
\vspace{-10px}
\end{figure}

This section focuses on obtaining the individual visits to a ROI and isolating them into the constituent components. 
Following the preprocessing steps to obtain the de-noised signal described in Section~\ref{sec:mobsig} and depicted in Figure~\ref{fig:preprocess}, we perform two more operations to make the peaks in the signals distinct. 
Note that, a visit is synonymous to a peak (upward or downward going) in the signal.

The first three steps involving the low-pass filtering, interpolation and the generation of a discrete signal are already discussed in Section~\ref{sec:mobsig}.  
The step after discretizing the location traces is curve fitting. 
As the peak shapes are not identical throughout the signal, a predefined, shape-dependent curve fitting could not be used to fit a curve to the cell IDs.
We observed that the peak shapes differ according to the visited place. 
For example, the movement is restricted to a relatively small area in a work/home/gym place, whereas it is dispersed over a larger area in a shopping mall or in a park. 
This does not affect the transition component of the peak, rather the cap (local static signal) of the peak representing the actual region.    
The peak shapes observed in the considered datasets can be represented in terms of convolution functions, i.e., $rectangular \ast Gaussian$ or $rectangular \ast Lorentzian$ or $triangular \ast Gaussian$ function.    
We do not make any assumptions regarding the shapes and perform a non-linear iterative-curve fitting with selectable peak-shape models.  
The curve fitting is applied to the whole signal, ensuring that the actual peak parameters are not distorted during the subsequent processing steps.
This step is crucial as it facilitates measurement of the slope to isolate the peak into its components (ROI+representative path).  
Furthermore, it is also necessary to accurately estimate the ROI area and the visit duration.  
The iterative fitting ensures that the peaks do not shift or are missed, which might result in inaccurate cell ID retrieval of user movements.

An elementary technique for peak detection is to take the first differential of the points whose peaks have a downward going zero-crossing at the peak maximum. 
However, the peaks can also lie below the basecamp that can be viewed as valleys. 
In this case, the upward-going zero-crossings are checked, and the local minima is accounted for, instead of the maxima.
The presence of white noise might result in false positives, leading to failure in obtaining the correct ROIs and estimating accurately the repeated visits.
It can also alter the derived features of the peak.
To address this, we apply a mean filter and smooth the first derivative prior to checking for the upward/downward-going zero-crossings.
Smoothing and differentiation can result in degrading the signal-to-noise ratio, which disturbs the peak shape and hence the peak entry and end times.
This is addressed by comparing the successive peaks against the previous peaks, assuming that no two peaks will be overlapped or directly adjacent to one another.
This assumption is valid because a user cannot be physically present at two distinct ROIs at any given time, and there must be a sufficient time gap (travel duration) between two successive ROI visits.

\begin{figure}[t]
\centering
\includegraphics[scale=0.58]{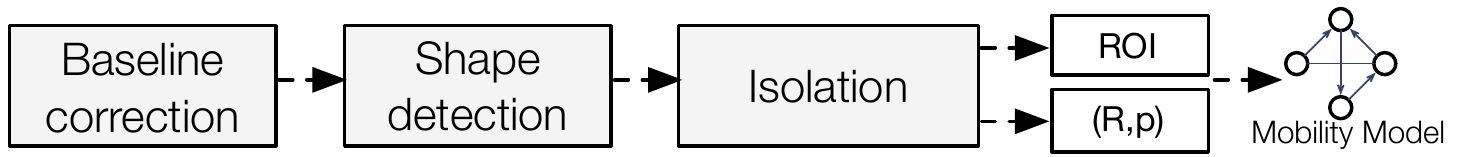}
\caption{Visit-detection and mobility modeling procedure.}
\label{fig:visit_detect}
\vspace{-15px}
\end{figure} 

\begin{figure}[t!]
	\centering
	\includegraphics[scale=0.23]{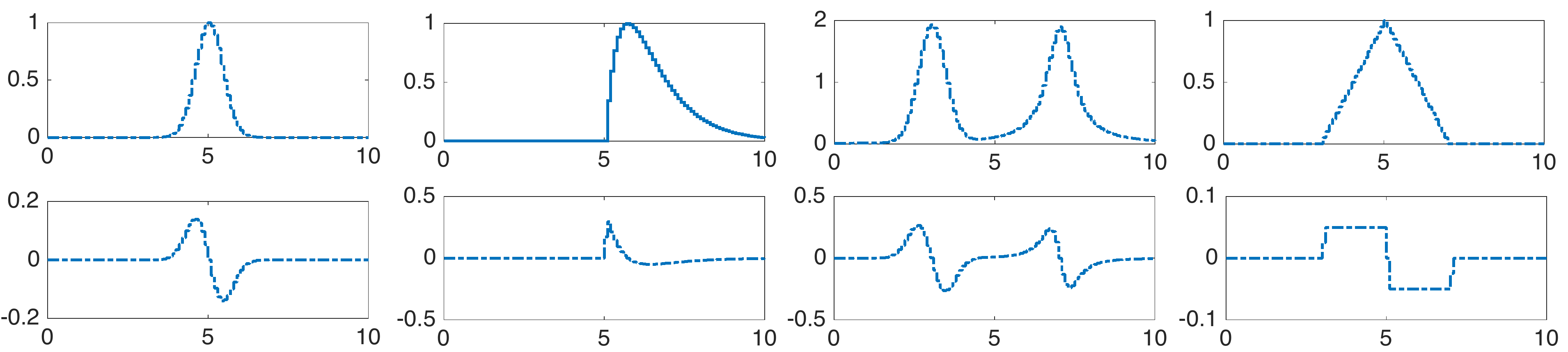}
	\caption{Peak Shapes (top row) and their respective derivatives (bottom).}
	\label{fig:derivatives}
\vspace{-15px}		
\end{figure}

Next, we pass the signal through the visit-detection and mobility modeling module depicted in Figure~\ref{fig:visit_detect}. 
Here, we perform the baseline correction, peak-shape detection and isolation. 
The baseline correction is performed to remove the background noise and to make the peaks distinct.
An important question is: How to automatically adjust the baselines so as to adapt constantly to changing user behaviors?
As our goal is to estimate the ROIs without mobility parameters, we do not perform flat-or quadratic-baseline corrections because such methods assume a complete view of the signal. 
To address this, we keep track of the standard deviation of the incoming points and analyze the points that deviate from the moving mean and the previous degree of standard deviation.  
This sets the baseline that works irrespective of the peak shape.
Furthermore, we need to correctly determine the peak shapes to accurately estimate the location, distance and time spent at the ROI.
The shape of the peaks can be detected by taking the successive derivatives, as different peak shapes have distinct derivative shapes as shown in Figure~\ref{fig:derivatives}. 

For example, a rising signal has a positive derivative, a signal that slopes down has a negative derivative, and a flat signal has a derivative that is zero. 
For the peaks associated with human movements, the accidence point coincides with the maximum of the first derivative, and it corresponds with the zero crossing point in the second derivative.
If Equation~\ref{eq:peak_detect} is satisfied, we consider the signal as a peak.

\begin{equation}
	\frac{d(S(t+1)-S(t))}{d(t)} - \frac{d(S(t)-S(t-1))}{d(t)} > 0
	\label{eq:peak_detect}
\end{equation} 

This process is not precisely instantaneous as we miss the peak by one $d(t)$, but this delayed detection compensates for false positives in the noisy data. 
Each $Visit$ can be separated into its constituent components by monitoring the average rate of change of slope.  
Upon arriving at a ROI, either the slope changes to zero or to an infinitesimally small value, as compared to the slope associated with the transition path component for some arbitrary slope $m \in \mathbb{R}$. 
The two parts are separable, depending on the average rate of change of the slope along the maxima or minima, such that~$Tr = Visit \mid \frac{\delta S(t)}{\delta t} = m$ and~$ROI = Visit \mid \frac{\delta S(t)}{\delta t} \neq m$. 
Once the cells belonging to the ROI are extracted, the remaining cells of the visit belong to the rising edge and the falling edge. 
In order to construct the representative path connecting the ROI, we rely on~\cite{Chapuis2016CapturingCB}. 
Our technique captures the practical nature of human mobility, by considering the fact that, users can move between two ROIs through different paths. 
We finally extract the best possible path amongst several options to represent the most significant trajectory of the user.

\begin{figure}[h!]
	\centering
	\includegraphics[scale=0.21]{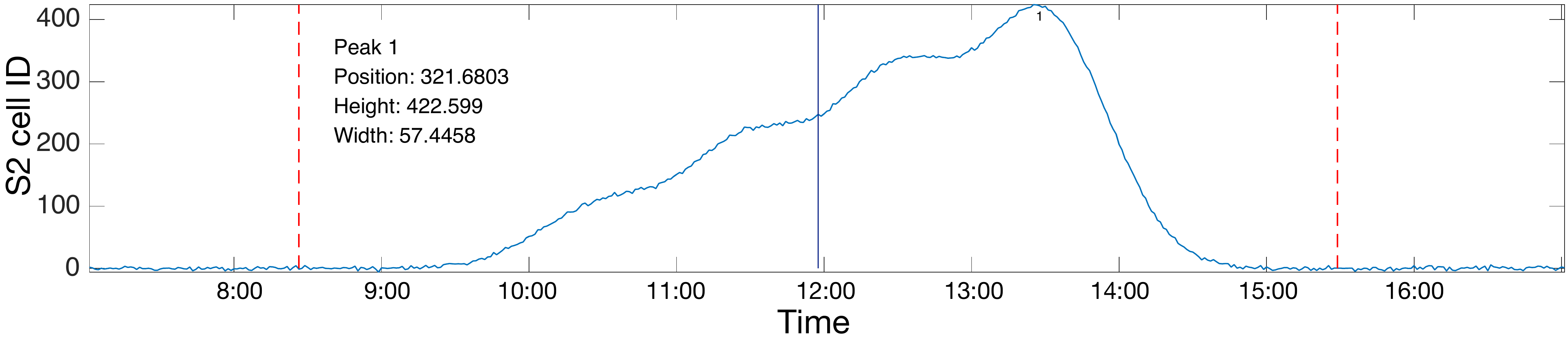}
	\caption{Peak detection and peak detail computation.}
	\label{fig:F2}
	\vspace{-10px}
\end{figure}

The positions where the slope changes also identifies the ROI entry and exit instants and are used to compute the area.  
Computing the zero-crossing in the first derivative gives the signal peak-point, irrespective of the signal type, hence the location of this point is an estimation of the mean visit time of the ROI. 
We attain the values of peak start and end in the process of peak-shape detection, i.e, the first derivative detects the time of the peak start and the second derivative gives the time of peak end as depicted in Figure~\ref{fig:F2}.  
This process finds the cells corresponding to the individual ROI of a user and the cells associated with the transition path component.

\subsection{Sub-ROI Discovery}

If the frequently visited region of a user is large, it might consist of a combination of smaller ROIs that we term as Sub-ROIs. 
For example, a university can be dissected into smaller regions such as the math building, engineering building and the cafeteria.  
Unlike the clustering techniques that require hierarchical clustering to dissect an extracted ROI to find these smaller locations~\cite{Ashbrook2003UsingGT}, we follow a different approach. 
A ROI can be visited by a user in two ways. 
Either the user follows a regular routine of visiting each Sub-ROI (e.g., math building to cafeteria) included in the main ROI (e.g., university) or simply visits one of the sub-ROIs (math building) and returns. 
A challenge here is to not characterize them as two different ROI visits, when both are essentially part of the same bigger ROI (university).

\begin{figure}[t!]
	\centering
	\includegraphics[scale=0.21]{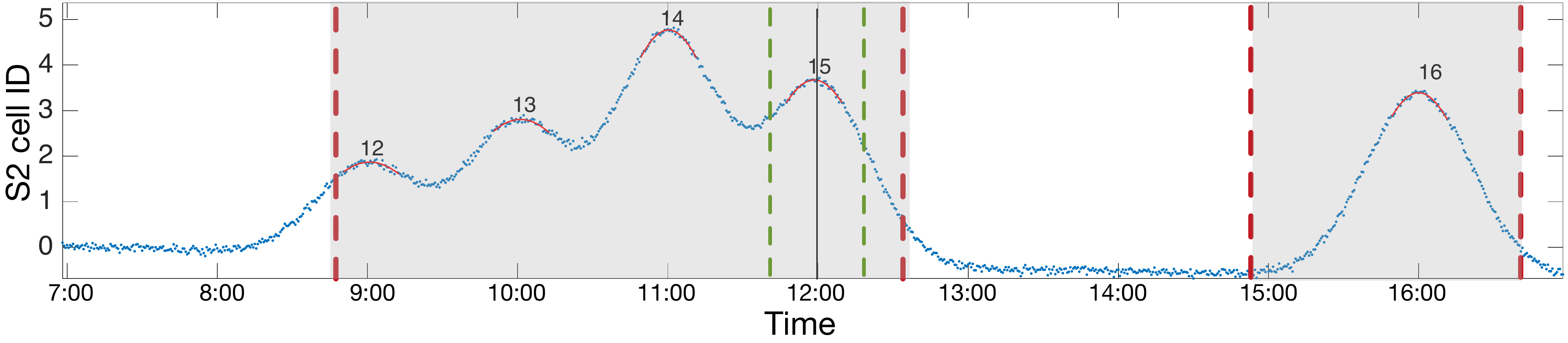}
	\caption{Repeated visit with a changed behavior creates a new ROI.}
	\label{fig:hotspot_rep}
\vspace{-10px}	
\end{figure}

A solution to address this issue is to check if a new peak/valley lies within the time frame of an already commenced peak/valley. 
The time bounds can be used to classify the minor peaks as a part of the major. 
However, if the behavior of the user changes as to take a different start and end route, the peaks can be classified as a new ROI, as shown in Figure~\ref{fig:hotspot_rep}.

To address this, we check if either of the peak-start and peak-end positions match, and we use Dice coefficient to check the similarity of the ROIs.
The Dice's coefficient can be expressed as in Equation~\ref{eq:dice_coffecient} and represents the similarity measure of two sets in range $[0,1]$, where 0 indicates no overlap.

\begin{equation}
	QS  = \frac{2|A\cap B|}{|A|+|B|}
	\label{eq:dice_coffecient}
\end{equation}

Here, A is the set of cells contained in the main ROI and B is the set containing the cells associated with the sub-ROI. 
Two ROIs cannot share the same cell, hence any value of $QS$ greater than 0 indicates an overlap.
This ensures that we classify the different patterns of visits in a ROI as repeated visits to the same.
A mobility model is thus formulated by linking all the distinct ROI's, the representative paths and the transition probabilities.
The transition probabilities are estimated based on a mobility Markov chain (MMC) model which accounts for the state-transition matrix as described in~\cite{Kulkarni:2016:MMP:3003421.3003424}.

\section{The Parameter Curse}
\label{sec:a-priori}

A key advantage of {\textbf{Capstone}} is the independence from the {\it{a–priori}} selected user behavioral parameters. 
To highlight this advantage, we demonstrate the bias and the privacy leakage resulting from the parameter space.
Given a user's spatial trajectory, the ROI discovery problem in the conventional setting is formalized as finding all the distinct ROIs, where each $ROI_i$ is a three-item tuple $(lat_i,lon_i,r_i)$, where $r_i$ is the radius of the region (assuming circular ROIs).  
Here, the maximum distance and the minimum time between two spatiotemporal points are bounded by fixed thresholds before assigning them to a particular cluster.  
These clusters are then merged, depending on their spatiotemporal similarity to form a ROI. 

\begin{figure}[t]
    \centering
    \begin{subfigure}[b]{0.23\textwidth}
        \includegraphics[width=\textwidth]{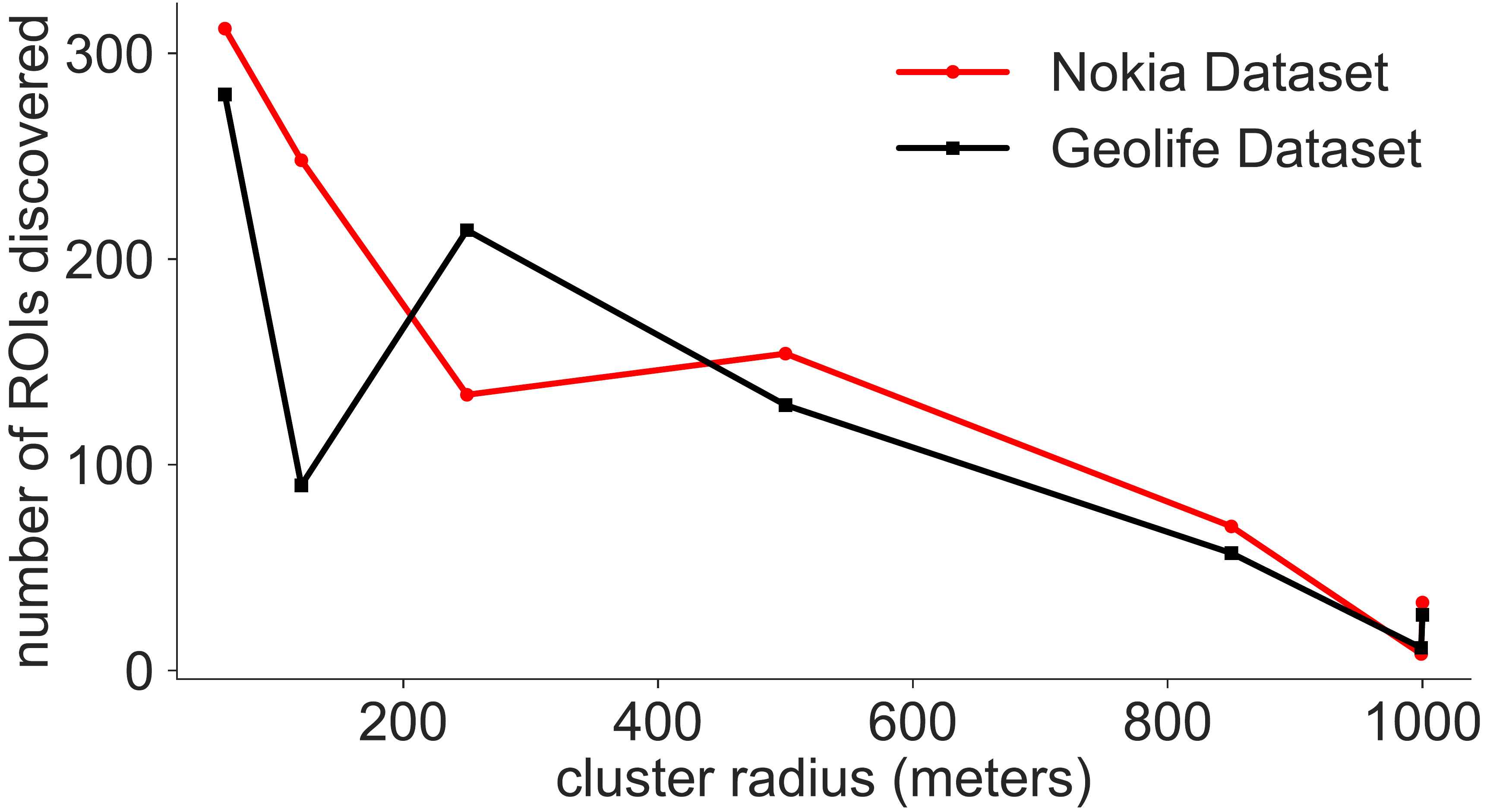}
        \caption{Number of ROIs vs. radius}
        \label{fig:cls_rad1}
    \end{subfigure}
    \begin{subfigure}[b]{0.23\textwidth}
        \includegraphics[width=\textwidth]{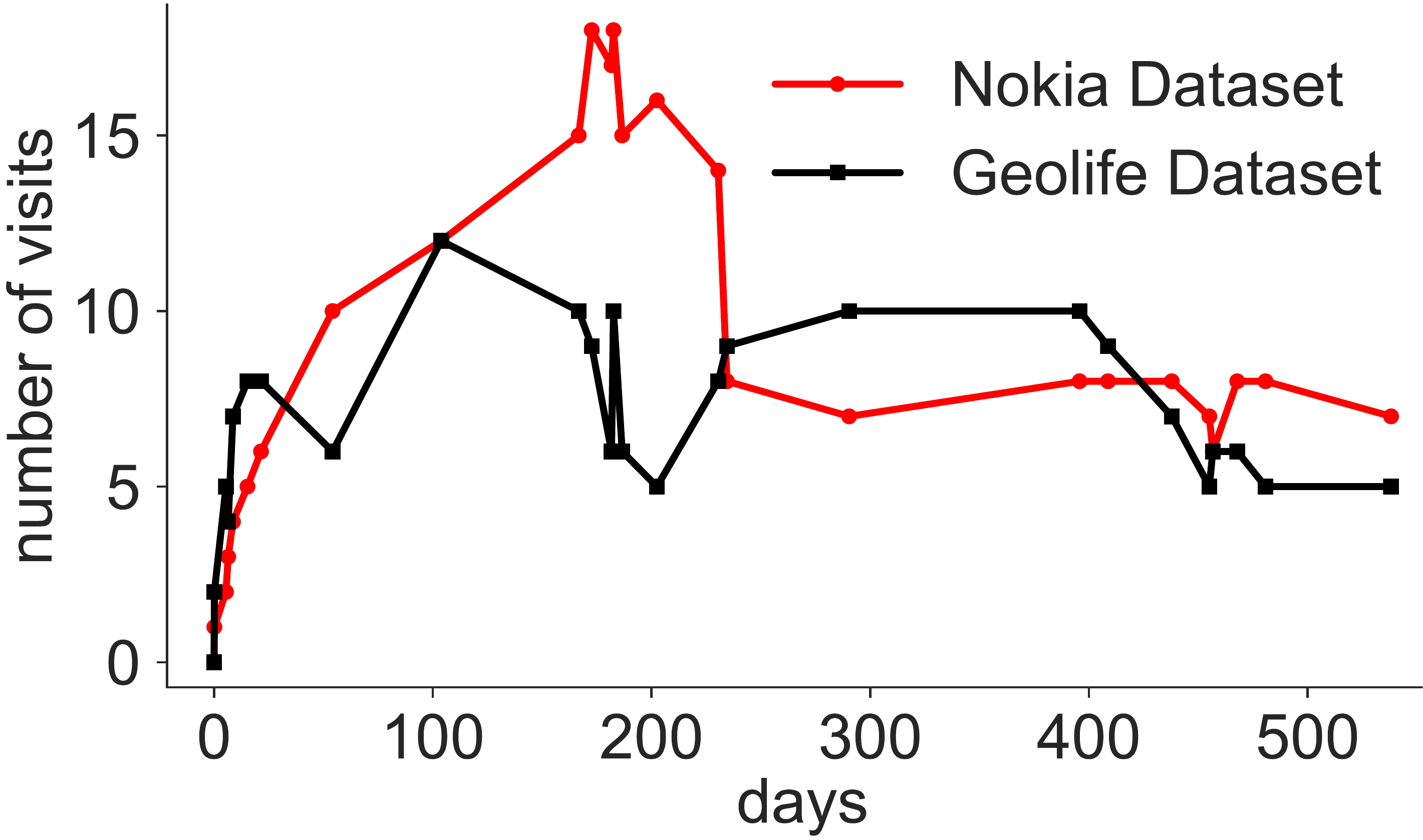}
        \caption{Number of visits vs. time}
        \label{fig:cls_rad2}
    \end{subfigure}
    	\caption{Trends across two different datasets for parameter estimation.}
    \label{fig:s}
\vspace{-15px}    
\end{figure}

In general, users are characterized by distinct mobility profiles, which results in different optimal values of these parameters. 
Their estimation is also challenging due to the large number of possible combinations, the duration of the available dataset, the sampling rate of the locations and the noise distribution in the recorded data.  
The parameter values are selected by analyzing the trends in the considered dataset or are derived by logical reasoning, which lacks exhaustive empirical basis.    
For example, the techniques that use cluster radius as a parameter~\cite{2ashbrook2002learning}, select it at a point, which results in a 'significant' change in the slope between the number of clusters vs. the cluster radii, known as a {\it{knee}} in the plot.  
We use the same technique proposed in~\cite{2ashbrook2002learning} to extract clusters and then derive the parameter values across two geospatial datasets as shown in Figure~\ref{fig:cls_rad1}. 
To select the value of {\it{minimum visit}}, either the knee in the plot of time vs. number of visits to a particular ROI is selected~\cite{Kulkarni:2016:MMP:3003421.3003424} as shown in Figure~\ref{fig:cls_rad2}, or the duration of the collected dataset is taken into account~\cite{3farrahi2011discovering}.  
We clearly see different trends followed by the two datasets.   
This results in a possible bias when generalizing and comparing the results obtained with different techniques or the same technique on different geospatial datasets.

Therefore, a comparison performed with a partial knowledge of the effect of altering the settings of the clustering algorithm will result in arbitrary conclusions. 
This has led to different parameter values in the published works that are dependent either on the application scenarios as in~\cite{4montoliu2010discovering}, or on the behavior as stated in~\cite{2ashbrook2002learning}, which results in inconsistent derivations of comparative results.
We also notice their disagreement about the significance of the effects of certain parameters.  
For example, the results highlighted in~\cite{4montoliu2010discovering} and~\cite{9thomason2016identifying} lead to conflicting conclusions regarding the importance of the {\it{maximum time}}, between two coordinate points.

Finally, these techniques extract clusters, characterized by fixed shapes (mostly circular).
In the literature, the circular cluster shape is based on the diffusion theory in Kulldorff's spatial scan statistics~\cite{Jung2010ASS}.   
The techniques that assume a pre-defined shape provide assurance of a complete enumeration of all the regions of that shape in a given area. 
However, clustering techniques, when applied for ROI detection assume a circular shape that might not represent reality. 
Setting predefined circular windows to define the potential cluster areas will result in difficulties in correctly detecting actual noncircular ROIs~\cite{Tango2005AFS} and furthermore to estimate the areas and total time spent. 
These issues show the importance of not relying on prior assumptions regarding either the parameters or shapes for devising generalizable and conclusive ROI-detection algorithms.
We present an exhaustive list of predetermined behavior dependent parameters used by popular ROI discovery techniques in Table~\ref{tbl:ROI_parameters}.

\begin{table}[t!]
\centering
\resizebox{\columnwidth}{!}{
\begin{tabular}{l c|l c|l c}
\hline
Parameter & Abbreviation & Parameter & Abbreviation & Parameter & Abbreviation \\ \hline
1. Max. Distance & $Max_{dist}$ & 6. Max. Point Separation & $Max_{dp}$ & 11. Grid Size & $S_{g}$ \\
2. Min. Time & $Min_{time}$ & 7. Gradient Threshold & $Tsh_{g}$ & 12. P Value & $Val_{p}$ \\
3. Max. Time & $Max_{time}$ & 8. Seed number & $Num_{s}$ & 13. Height & H\\
4. Num. Eigenvectors & $Num_{ev}$ & 9. Vector Length & $Len_{vec}$ & 14. Cluster Radius & $C_r$\\
5. Min. Num. Points & $Min_{points}$ & 10. Minimum Visits & $Min_{visit}$ & 15. Minimum Speed & $Min_{speed}$\\ \multicolumn{1}{l}{} & \multicolumn{1}{l}{} \\ \hline
\end{tabular}}
\caption{Parameters used by existing ROI discovery techniques.}
\label{tbl:ROI_parameters}
\vspace{-15px}
\end{table}

These parameters are measures of individual mobility dynamics~\cite{Pellungrini2017ADM}. 
Therefore, in a situation where the adversary requests a data provider for aggregated/sparse mobility data, a knowledge of these parameters can increase the background knowledge to carry about membership inference attacks. 
Parameters such as radius of gyration, mobility entropy and average number of visits of an individual, have been shown to de-anonymize users from aggregated databases~\cite{xu2017trajectory}. 
A recent work to estimate privacy risk of individuals based on the individual mobility features show that several parameters such as the maximum distance/time between locations, total distance traversed per day, number of distinct locations and others increase an individual's risk of identification against location sequence construction attacks, home and work place attacks, location probability attacks, etc.~\cite{Pellungrini2017ADM}. 
Therefore, we argue that a technique to extract user mobility models even from sparse data without relying on user's mobility parameters is beneficial.

\section{Evaluation and Discussion}
\label{sec:evaluation}

In this section, we evaluate {\textbf{Capstone's}} effectiveness in mobility modeling without any parameters and its operational efficiency on smartphones based on the implementation of the proposed technique on a DSP chip.
We also perform privacy analysis, by quantifying the accuracy and risk of two popular attacks performed on the user's exposed locations.   
All the evaluations are performed using the Nokia dataset~\cite{Nokiadataset1}, Geolife~\cite{zheng2010geolife} and a third dataset annotated with the ground truth. 
These datasets contain geospatial trajectories of more than 370~users, collected in Switzerland and China. 
ROIs being the key component of the mobility model, our focus is on their validation.  
The extraction of representative trajectories from the set of transition paths, provides precision and recall rates exceeding 80\% as shown in~\cite{Chapuis2016CapturingCB}.

We configure the google S2 library to project each coordinate pair onto a cell of dimension $38m^2$. 
It could be argued that the cell size involves a arbitrarily chosen parameter in the process. 
However, our choice is motivated by the localization accuracy of a typical GPS sensor and the performance complexity involved when subdividing the cells to the leaf level.

The publicly available datasets are devoid of the ground truth due to privacy concerns.   
Therefore, we collect an additional dataset by providing a mobile application to one of the co-authors of this paper as a part of our data collection campaign.{\footnote{Mobility data collection campaign: bread-crumb.github.io}}
The application logs the latitude, longitude, timestamp, acceleration, altitude, horizontal and vertical accuracy of the GPS coordinates.  
The data points are collected at a sampling rate of 5 seconds with a granularity of resolution up to 5 meters for a period of 15 weeks. 
The ground truth is captured by periodically attesting the visited regions of interest, average time spent and the approximate area. 

\begin{figure}[t!]
    \centering
    \begin{subfigure}[b]{0.241\textwidth}
        \includegraphics[width=\textwidth]{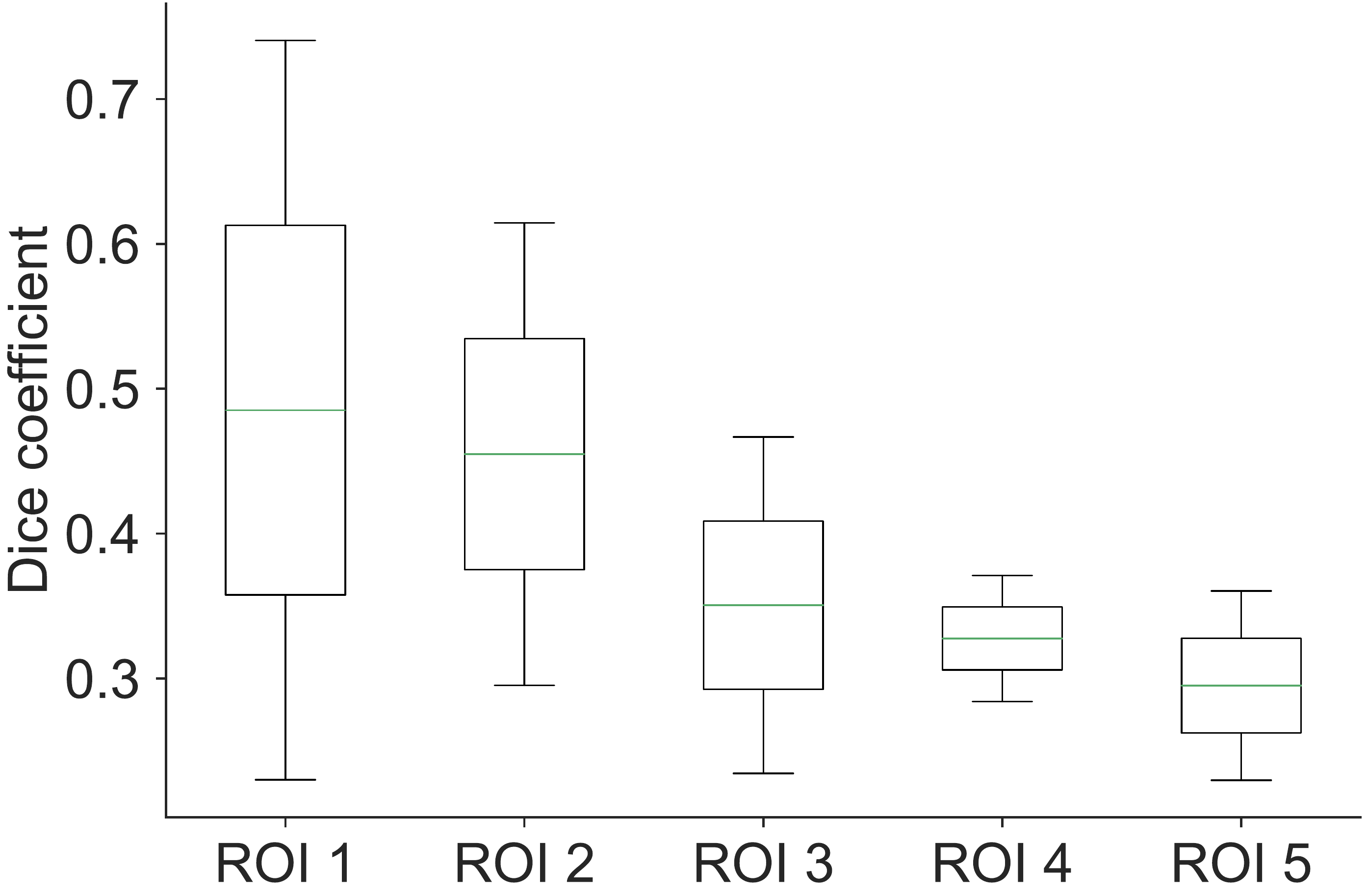}
        \caption{Dice index distribution}
        \label{fig:dice2}
    \end{subfigure}
    \begin{subfigure}[b]{0.241\textwidth}
        \includegraphics[width=\textwidth]{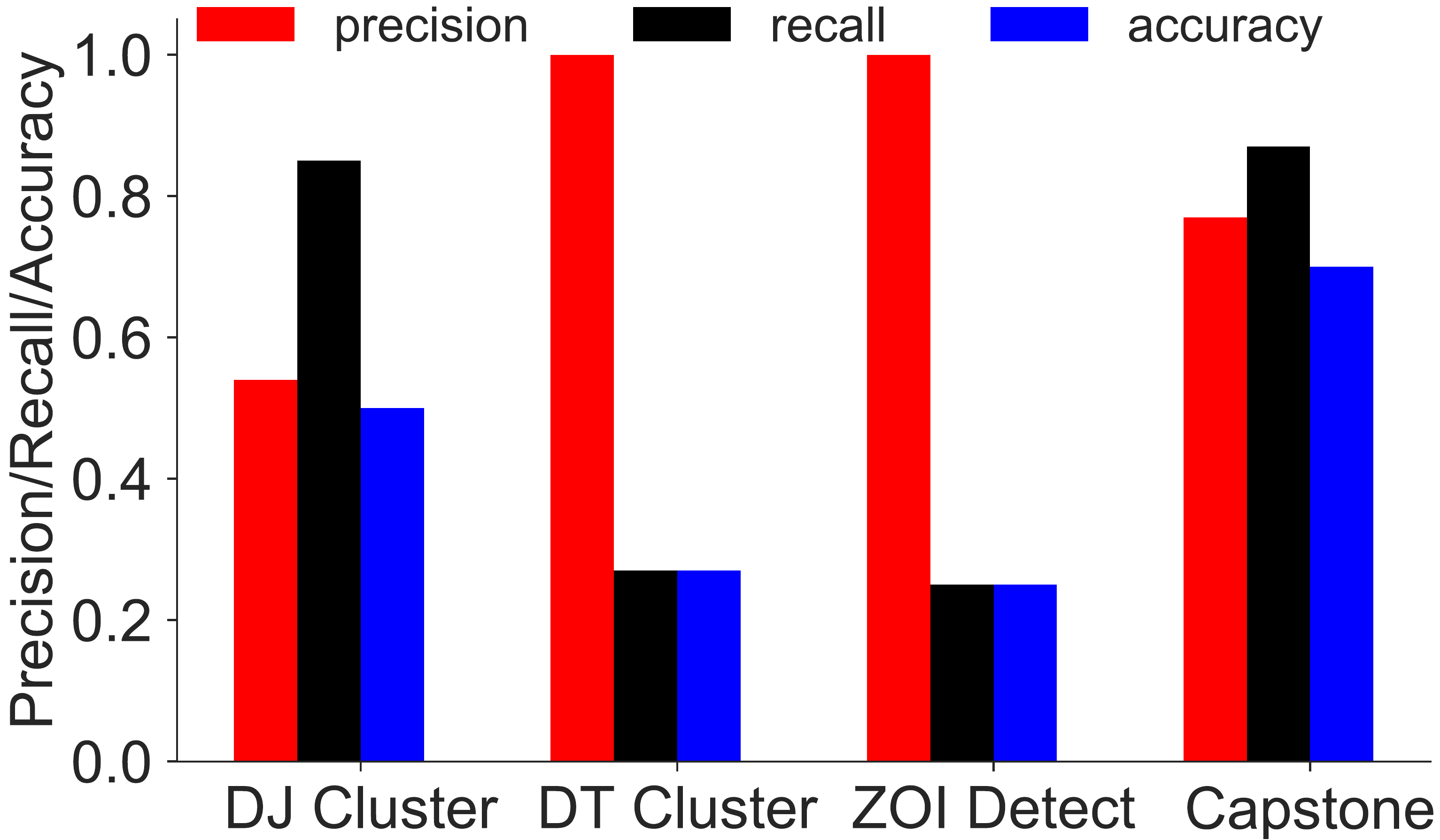}
        \caption{Ground truth validation}
        \label{fig:ground_truth}
    \end{subfigure}   
    \caption{Comparison of the ground truth analysis.}
    \label{fig:comp1}
\vspace{-15px}    
\end{figure}

\subsection{Visit Consistency}

Here, we perform a qualitative evaluation by using the mobility datasets to guarantee consistency of the discovered ROIs using metrics derived in published works based on the same datasets~\cite{9thomason2016identifying}.

\begin{table}[h]
\centering
\resizebox{\columnwidth}{!}{
\begin{tabular}{c|ccc|c|ccc}
\hline
\textbf{Num. ROIs} & 2-5 & 6-9 & 10-12 & \textbf{Max. Stay Time} & 5-8 Hrs & 9-10 Hrs & 11-26 Hrs \\
\textbf{User \%} & 37 & 51 & 12 & \textbf{User \%} & 67 & 26 & 7 \\ \hline
\textbf{Short Visits} & \textless10 min & \textless15 min & \textless30 min & \textbf{Stay Time v/s Travel time} & 3:2 & 4:1 & 2:3 \\
\textbf{User \%} & 22 & 36 & 42 & \textbf{User \%} & 35 & 54 & 11 \\ \hline
\end{tabular}}
\caption{Visit accuracy evaluation based on Nokia dataset.}
\label{tb:dataset_summary}
\vspace{-5px}
\end{table}

The task of ROI extraction is synonymous with unsupervised clustering. 
Therefore, we first validate our results by relying on the knowledge about the data and the properties of ROI visits.
To this end, we use the properties derived by Thomason et al.~\cite{9thomason2016identifying}, which hold for a majority of the Nokia dataset users. 
They comprise of: (i) A typical user makes an average of 2 to 15 distinct ROI visits a day, (ii) a visit does not exceed a period of 2 days, (3) a user spends 60\% of the time at the ROI and not more than 40\% traveling to the location. 
Our results (see Table~\ref{tb:dataset_summary}) corroborate these properties.

A drawback of our approach is its high sensitivity to even small stoppages occurring on the path, due to which unintentional delays can be classified as ROIs, e.g., a bus stop on the way of an intentional destination. 
Furthermore, we do not rely on the $Min_{visit}$ parameter, which classifies even a single visit as a ROI and results in some false positives. 
This parameter is often selected depending on the duration of the available dataset, which does not reflect the true periodicity of visiting a particular place.   
If the periodicity is very low, this will be reflected through the learning algorithms if the extracted ROIs are utilized for applications such as mobility prediction~\cite{CABOU04}. 
Thus, although our technique might result in some false positives, it ensures that none of the ROIs are filtered out either based on the dataset duration or the mobility behavior. 
The additional outliers occur due to property (iii) which does not hold for bus/metro/train stops.  

Next, we examine if the repeated ROI visits with a differing user behavior (different entry or/and exit points) results in creation of new ROIs. 
To analyze this, we use the Dice coefficient to assess the similarity, and show its distribution across the considered set of ROIs spanning distinct areas.
As depicted in Figure~\ref{fig:dice2}, marginal deviations in user behavior associated with repeated visits, does not lead to creation of new ROIs.

\subsection{ROI Accuracy}
\label{subsec:roi_acc}

In this section, we perform quantitative analysis using the dataset annotated with the ground truth and compare our results with three popular time-space-density-based clustering algorithms commonly used for ROI extraction.

\begin{table}[t!]
    \centering
    \resizebox{\columnwidth}{!}{
    \begin{tabular}{c c}
    \toprule
    Clustering algorithm	 & Parameters\\
    \midrule
    DJ Cluster & $Min_{speed}$: 0.4 (km/hour) / $C_r$: 60.0 (meters) / $Min_{points}$: 10\\
   	DT Cluster & $Max_{dist}$: 60.0 (meters) / $Min_{time}$: 900 (seconds)\\
    ZOI Detect & $Max_{dist}$: 60.0 (meters) / $Min_{time}$: 900 (seconds) / $Min_{visit}$: 6\\
    \bottomrule
    \end{tabular}}
    \caption{Clustering algorithms with the default parameter values.}
    \label{tab:clustering_parameters}
\vspace{-10px}    
\end{table} 

\begin{figure*}[t!]
    \centering
    \begin{subfigure}[b]{0.30\textwidth}
        \includegraphics[width=\textwidth]{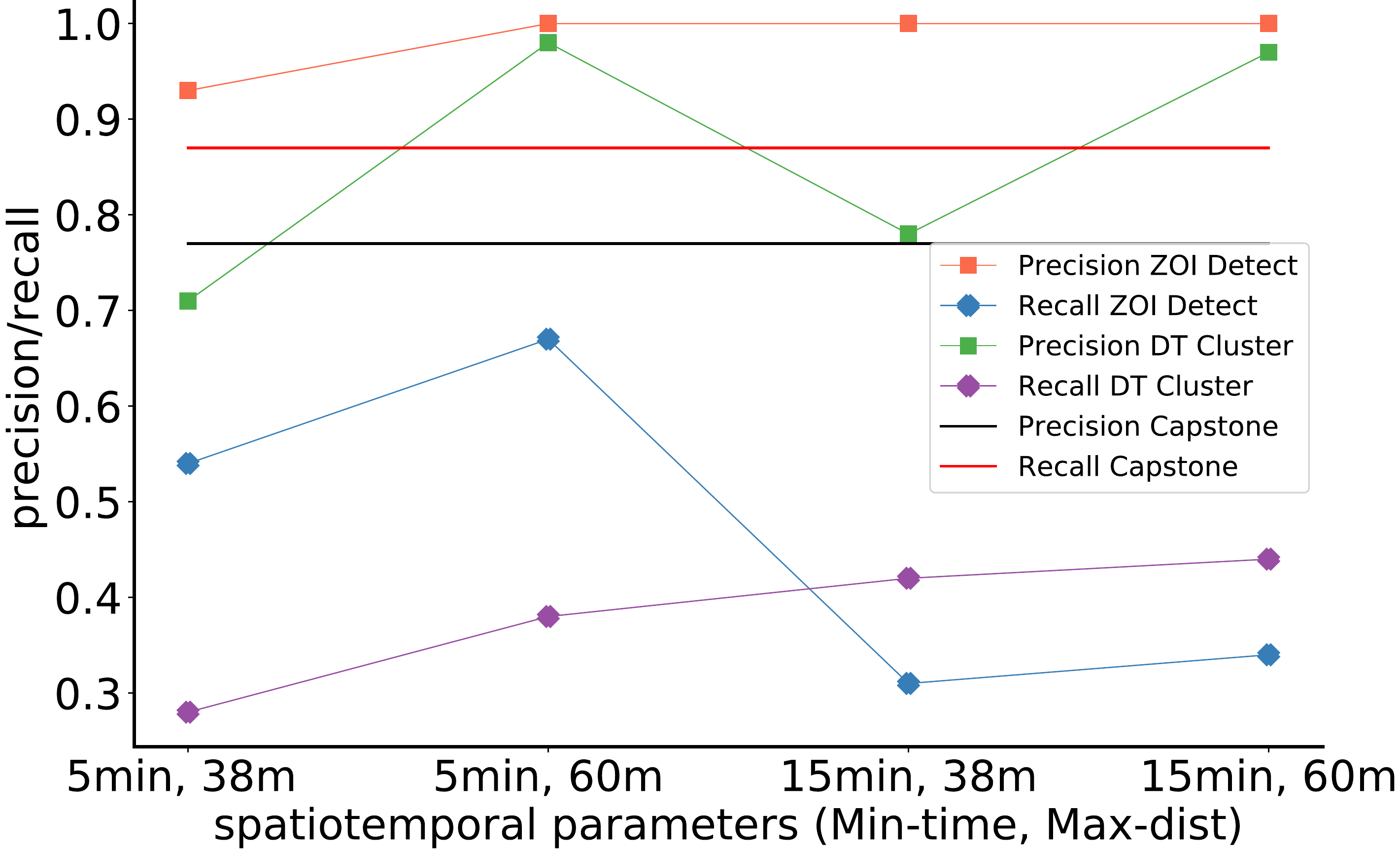}
        \caption{Ground truth validation}
        \label{fig:ground_truth_all}
    \end{subfigure}
    \begin{subfigure}[b]{0.34\textwidth}
        \includegraphics[width=\textwidth]{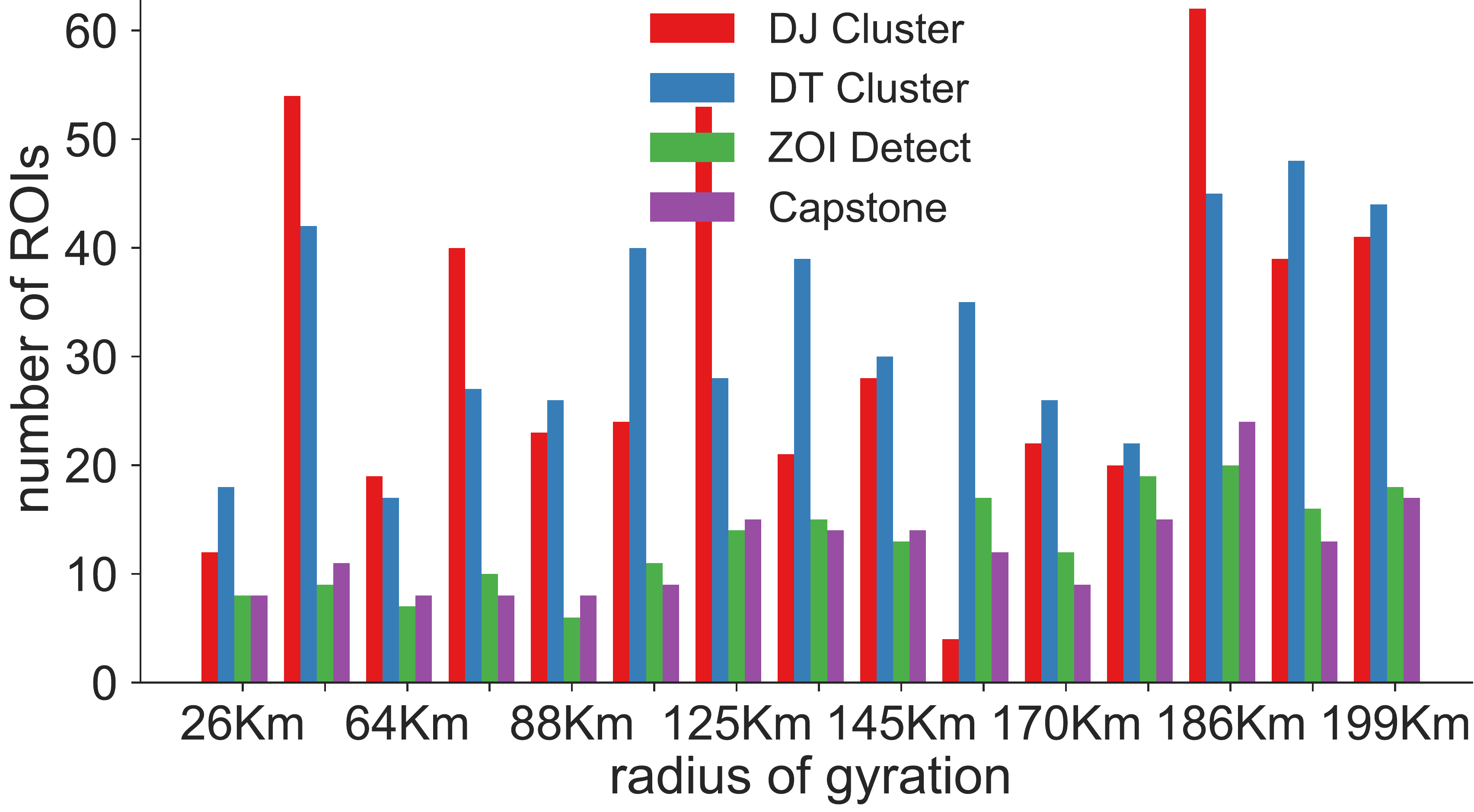}
        \caption{Number of ROIs: Nokia dataset}
        \label{fig:cluster_comp}
    \end{subfigure}
    \begin{subfigure}[b]{0.34\textwidth}
        \includegraphics[width=\textwidth]{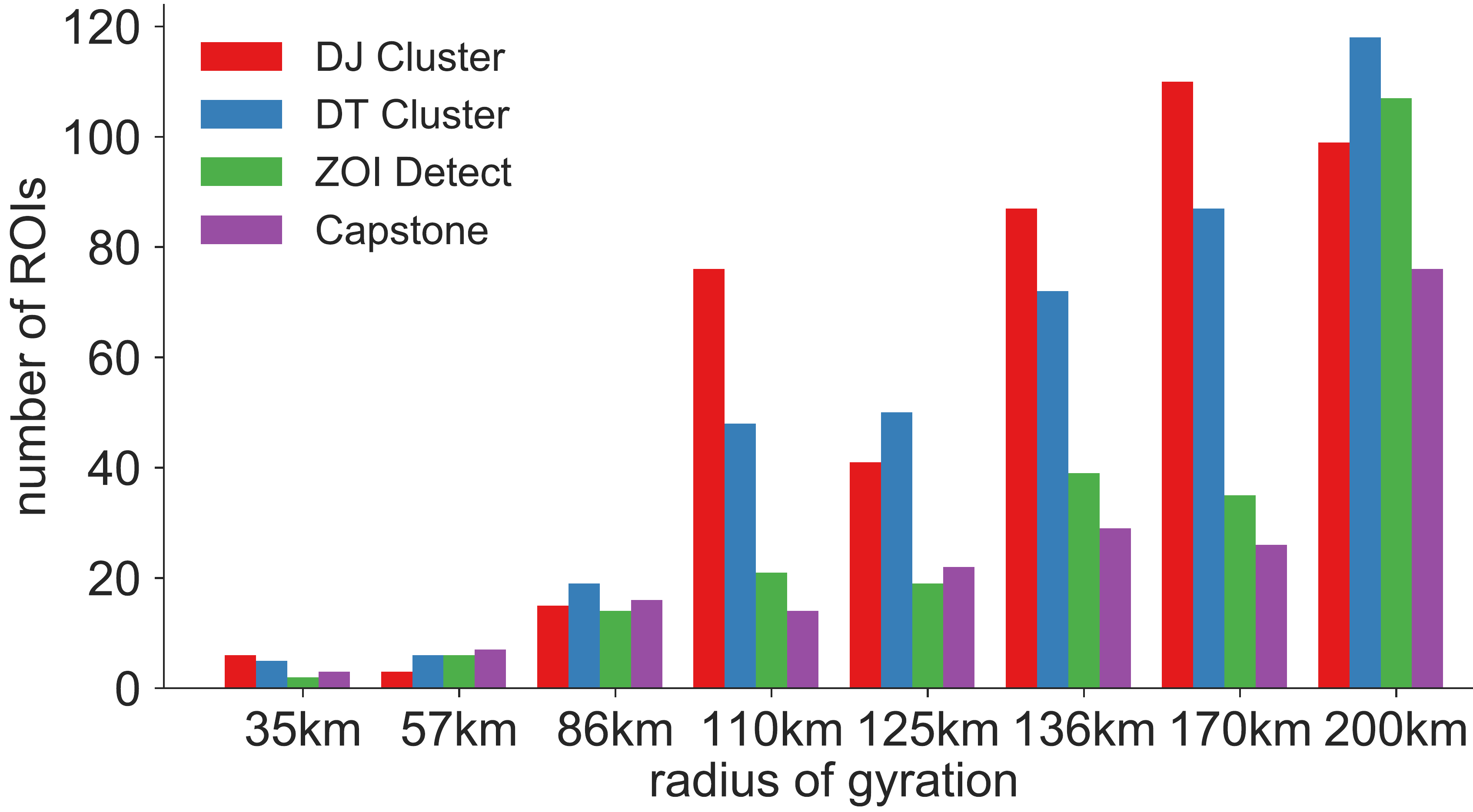}
        \caption{Number of ROIs: Geolife dataset}
        \label{fig:comp_geolife}
    \end{subfigure}   
    	\caption{Comparison of the ground truth with different parameter settings and the number of ROIs using two different datasets.}
    \label{fig:comp2}
\vspace{-5px}    
\end{figure*}

Here, we validate the accuracy of the discovered ROIs with respect to the ground truth and compare our results with three clustering techniques.  
We consider Density Joinable Cluster (DJ Cluster)~\cite{zhou2004discovering}, Density Time Cluster (DT Cluster)~\cite{hariharan2004project} and ZOI Detect~\cite{Kulkarni:2016:MMP:3003421.3003424}. 
DJ Cluster computes ROIs based on the number of points within a certain radius and merges the clusters if they share at least one common point. 
The points are also clustered together if they satisfy the $Min_{speed}$ bound.  
DT Cluster aggregates points lying within predetermined spatiotemporal bounds.
These clusters are then treated as valid ROIs.  
ZOI Detect follows a similar strategy as DT cluster but relies on an additional parameter $Min_{visit}$ as a threshold and merges the clusters upon intersection.    
The parameter space of these techniques and their values are shown in Table~\ref{tab:clustering_parameters}.
These values selected in published works are based either on the dataset trends~\cite{Kulkarni:2016:MMP:3003421.3003424} or on the mobility behaviors~\cite{zhou2004discovering}.

For the ground-truth comparison, the 34 ROIs of the considered subject were selected with a clear definition: 'any place where the subject visited with an intentional purpose'. 
These regions include places such as cafeterias, restaurants, bus/train/metro stops, sports arenas, bookstores, office and work places, and excursions.             
The ground-truth evaluation was performed by computing the precision, recall and accuracy. 
As the set of true negatives is infinite (ROIs not visited by the user and not discovered by the algorithm), $Accuracy = \frac{TP}{TP+FP+FN}$. 
A comparison of {\textbf{Capstone}} with the clustering techniques, with respect to the ground truth, is shown in Figure~\ref{fig:ground_truth}. 
To ensure consistency, the ROIs with the clustering techniques are computed with the default parameter values given in Table~\ref{tbl:ROI_parameters}.

We see that DT Cluster and ZOI Detect have a very high precision and low recall and accuracy.   
This indicates that these techniques detect a large number of ROIs that are not contained in the true ROI set. 
This is clearly due to the spatiotemporal bounds being too rigid, which results in considering arbitrary clusters as ROIs.  
DJ cluster, however, has higher recall and low precision. 
Here, we see that the $Min_{speed}$ eliminates the occurrences of false negatives, whereas, the $Min_{points}$ creates high number of false positives.
Increasing the $Min_{points}$ can address such occurrences, as it requires a higher density of points, thus creating only valid ROIs.   
In case of {\textbf{Capstone}}, we have a few false positives due to the high sensitivity and only three false negatives. 
The false negatives are the transportation stops where the user does not have to wait due to planned time synchronization, resulting in a constant average slope.

To better understand the parameter influence, we consider four different parameter sets for the values of $Min_{time}$ and $Max_{dist}$ as seen in Figure~\ref{fig:ground_truth_all}.
We see that the parameter $Min_{visit}$ always correctly classifies a region as a ROI, thus leading to high precision rates.  
We can also see that larger values of $Max_{dist}$ results in higher precision and recall in DT Cluster.  
$Max_{dist}$, thus plays a vital role in determining precision, compared to $Min_{time}$ parameter in the considered dataset. 
These results highlight the importance of selecting the parameter space which is challenging to determine a–priori.

To present qualitative results we evaluate the Nokia and Geolife dataset.
In the absence of ground truth, choosing relevant metrics for comparison is a challenging problem. 
To address this, we explore the number of ROIs discovered as it directly influences the accuracy of the technique. 
A lower number may signify the merging of multiple ROIs leading to the loss of information, such as the total area and the time of entry and exit from the respective ROI. 
Whereas, a large number indicates a higher number of false positives. 
We first show the results for the Nokia dataset in Figure~\ref{fig:cluster_comp} and the Geolife dataset in Figure~\ref{fig:comp_geolife}. 
In order to consider different mobility behaviors, we select users with distinct activity areas captured with respect to the radius of gyration of movement.

DJ Cluster and DT Cluster detect a significantly high number of ROIs, not typical for an average user.
In case of DJ Cluster, we find that the parameter $Min_{points}$ creates a large number of ROIs.
However, we argue that if the sampling rate of the dataset is high, the $Min_{speed}$ could play an important role in further increasing the number of clusters. 
Whereas, in DT Cluster $Min_{time}$ parameter results in a higher frequency of visit separations increasing the total number ROIs.
We see that the number of ROIs discovered by ZOI Detect is lower than DT and DJ Cluster.
This is due to the merging of individual clusters upon intersection, in addition to extracting the most frequent clusters governed by the $Min_{visit}$ parameter.   
In general, if the parameters satisfy cluster merging, multiple clusters merge and form a large ROI; ROI division occurs if this bound is missed by even an infinitesimal small value. 
This results in the fluctuation of the number of ROIs solely due to the parameters.    
We cannot validate the accuracy of the ROIs detected by {\textbf{Capstone}} in this case, however, we observe a consistency between the distance and the ROI number.
We also do not observe an alarming number of ROIs.  
We exclude DBSCAN~\cite{citeulike:3509601} and TD clustering~\cite{Gambs:2011:SMY:2019316.2019320} from the comparison, as they are similar in the parameter space to the techniques already considered.

The ROI area in our approach corresponds to $38m^2\times cell numbers$. 
This results in a significantly smaller areas compared to the clustering techniques and overlaps with the ground truth area. 
This is due to the set of cells comprising the ROI, which corresponds to the actual ROI area relative to the user movement.  
In contrast, the clusters encompass the area not covered by the user, due to reliance on bounding circles, where the centroid corresponds to the mean of the points and the radius corresponds to the maximum distance between the centroid and the points.
The same reasoning holds for the time spent in the ROI. 
This representation method can introduce a large area of dead space, and we argue that convex hulls would be more suitable than bounding circles for representing ROIs in the clustering techniques.

\subsection{Complexity and Power Consumption}

\begin{figure}[t]
\centering
\includegraphics[scale=0.4]{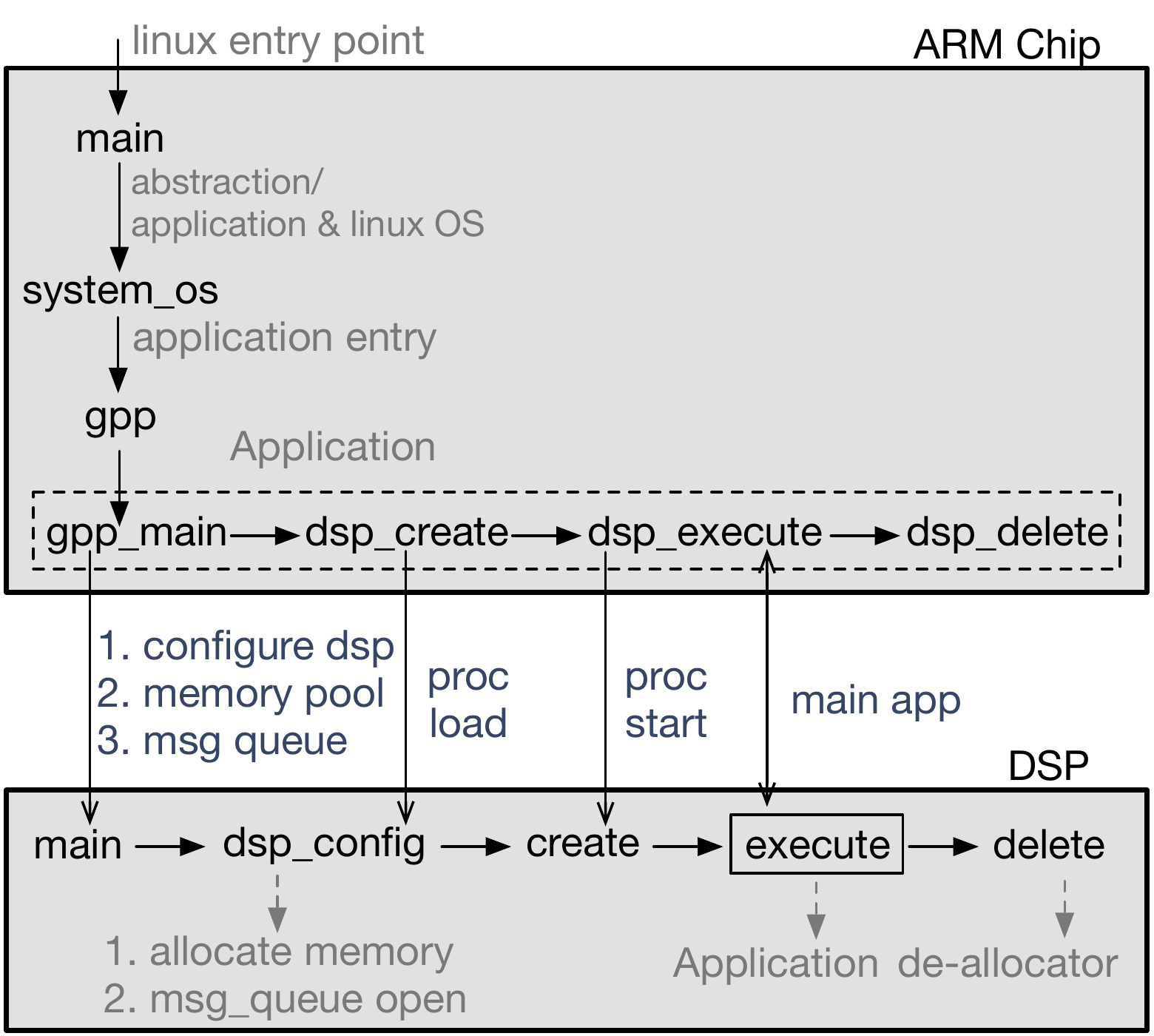}
\caption{A typical workflow of a GPP-ARM based SoC.}
\label{fig:arm_ggp}
\vspace{-15px}
\end{figure}

\begin{figure}[!t]
    \centering
    \begin{subfigure}[b]{0.195\textwidth}
        \includegraphics[width=\textwidth]{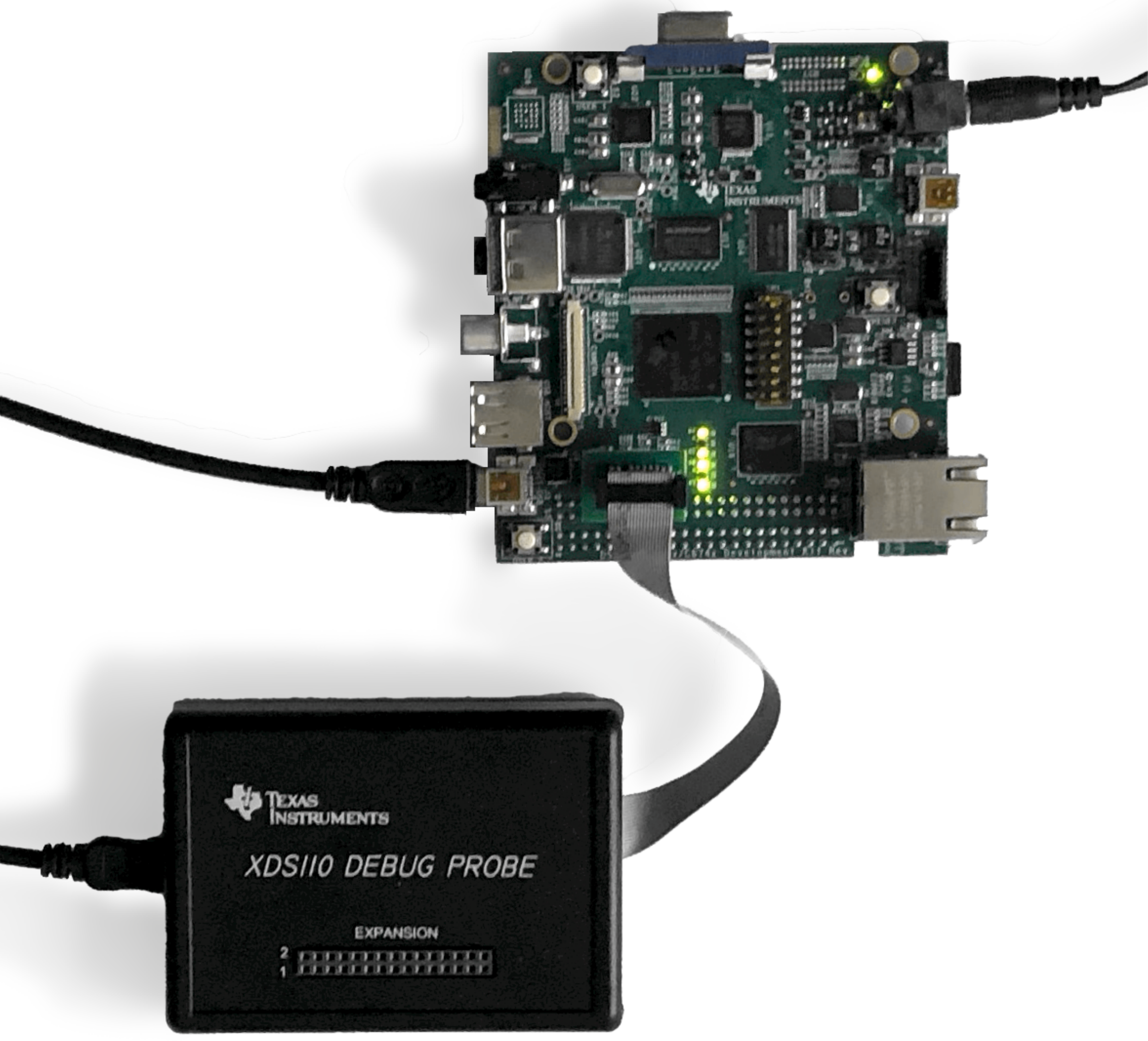}
        \caption{Experimental setup}
        \label{fig:exp_setup}
    \end{subfigure}
    \begin{subfigure}[b]{0.285\textwidth}
        \includegraphics[width=\textwidth]{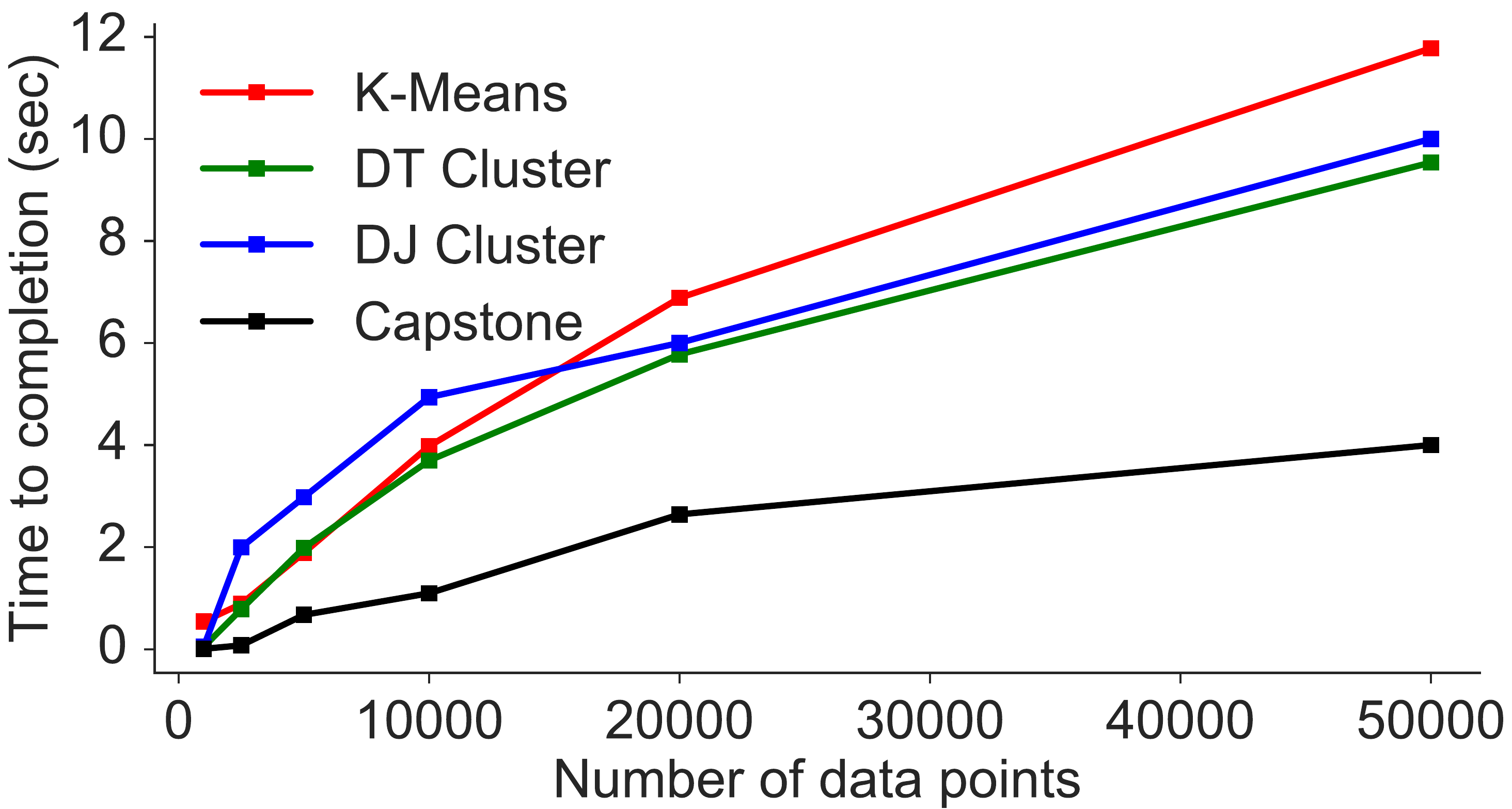}
        \caption{Execution time comparison}
        \label{fig:comp_plot}
    \end{subfigure}   
    \caption{Experimental setup and performance comparison.}
    \label{fig:exp_setup_comp}
\vspace{-15px}    
\end{figure}

Next, we evaluate and compare the techniques with respect to their computational complexity and power consumption. 
We implement Capstone, DJ Cluster, DT Cluster and K-means (as a reference) on a TI OMAP-L138 C6000 DSP+ARM Processor (Figure~\ref{fig:exp_setup}) present in many smartphones.{\footnote{TI C6000: www.ti.com/product/OMAP-L138}}
The Dual-Core SoC contains an ARM9 general purpose processor (GPP) and a C674x DSP core. 
As the performance and scaling is also dependent on the actual implementation of the algorithms, we do not optimize any techniques and derive only the asymptotic performance. 
A typical workflow between the GPP and the ARM processor is depicted in Figure~\ref{fig:arm_ggp}.

We benchmark the performance at various dataset sizes and consider the average time after 10 runs on each dataset size (see Figure~\ref{fig:comp_plot}). 
We see that, capstone reduces the runtime latency as compared to the rest by a factor of approximately 2.5.
The key reasoning behind the performance is: (i) the stackable non-blocking signal-processing pipelines, (ii) maximizing the computations per clock cycle by mapping these pipelines to the DSP architecture, (iii) efficient execution of all the filtering and peak-detection stages by utilizing the five multiply-add-accumulate units (MAC) in parallel, and (iv) space-transformation which facilitates carrying out all the operations on integers rather than 3-dimensional floating points. 
The performance of K-means rapidly deteriorates as the execution depends on the disk IO bound, and continually paging the RAM to access the distance array dramatically increases the runtime.    
Similarly, the agglomerative/hierarchical clustering techniques suffer through the same drawback.   
An additional drawback of such algorithms is due to the fact that they operate in several steps~\cite{2ashbrook2002learning}. 
This is done, by first clustering the points in the temporal domain and consequently in spatial domain, or by extracting locations that span large areas and dissecting them into smaller regions in the second iteration over the dataset. 
This results in increased time and computational complexity, hindering the possibility of operating them in real-time scenarios on resource-constrained devices.

In order to theoretically compute the complexity bounds, we consider a total of $n$ coordinate points from which the ROIs have to be extracted. 
We assume $k$ unknown ROIs, as we do not have a-priori knowledge on the number of clusters that will be detected. 
In the case of space-time-density based clustering techniques, there are multiple blocking steps involved.
For each coordinate assignment to a {\it{stay region}}, the Euclidian distance and time bounds are computed, and checked with the neighboring points.
Once the stay region is estimated, the centroid of the region is computed. 
This step has an overall complexity of $O(kn)$. 
The next step involves iterative merging of clusters based on distance bounds and is characterized by a complexity of $O(\sum_{i=0}^{k-1}(k-1)^2)$.
Scalar product of both the steps measures the total complexity. 
In case of Capstone, the preprocessing and peak-detection steps, the low-pass filtering, curve fitting, and the mean filtering contribute to a complexity of $O(2n)$; and the differentiation and baseline corrections contribute to $O((2n)^2)$, which results in a total complexity of $n^2$.
We can consider the operations as a $n\times n$ scalar matrix $C$ multiplying a scaler vector $v$ of length $n$; these operations result in a total of $n^2$ multiplications and $n(n-1)$ additions.
These multiplications and additions are parallely executed across the five MAC units in a non-blocking fashion contributing to the runtime improvement.


Next, we compare the power consumption at various dataset sizes as shown in Table~\ref{tbl:pwr_con}.
The power drawn by a process can be categorized in to baseline and active power. 
The former includes the static power (leakage), phase-locked loop, oscillator power and various subsystem components that cannot be turned off through the on-chip power management module. 
Active power is the consumption due to the active parts of the SoC, which is dependent upon the frequency, utilization, read/write balance and switching (GPP-DSP). 
We consider the total power as the sum of these individual power consumptions measured using the TI's EnergyTrace tool{\footnote{Energy Trace: www.ti.com/tool/ENERGYTRACE}}.    

\begin{table}[h!]
\centering
\resizebox{\columnwidth}{!}{
\begin{tabular}{l|l|l|l|l}
\hline
\# data points & 1000 & 10000 & 50000 & 100000 \\ \hline
K-Means & {\bf{0.27$mW$}} & {\bf{1.67$mW$}} & 6.76$mW$ & 11.93$mW$ \\
DT Cluster & 0.33$mW$ & 2.33$mW$ & 8.23$mW$ & 14.66$mW$ \\
DJ Cluster & 0.39$mW$ & 3.14$mW$ & 8.95$mW$ & 15.28$mW$ \\
Capstone & 3.13$mW$ & 3.78$mW$ & {\bf{4.04$mW$}} & {\bf{6.79$mW$}} \\ \hline
\end{tabular}}
\caption{Power consumption comparison (baseline + active power).}
\label{tbl:pwr_con}
\vspace{-5px}
\end{table}

{\textbf{Capstone}}; inherently a DSP implementation draws a higher baseline power as compared to the GPP implantation of the clustering techniques.
This is due to the power consumed in configuring the DSP chip and setting up the shared memory pool, the message queue between the GPP and the DSP and the real-time operating system (RTOS).
We clearly see in the results that, as the dataset size increases the power consumption of the clustering techniques rapidly escalates. 
However, the DSP implementation leverages the efficient power management capabilities of the RTOS that uses the chip power-efficiently, while still providing high performance.

\subsection{Privacy Analysis}

\begin{figure}[t!]
    \centering
    \begin{subfigure}[b]{0.24\textwidth}
        \includegraphics[width=\textwidth]{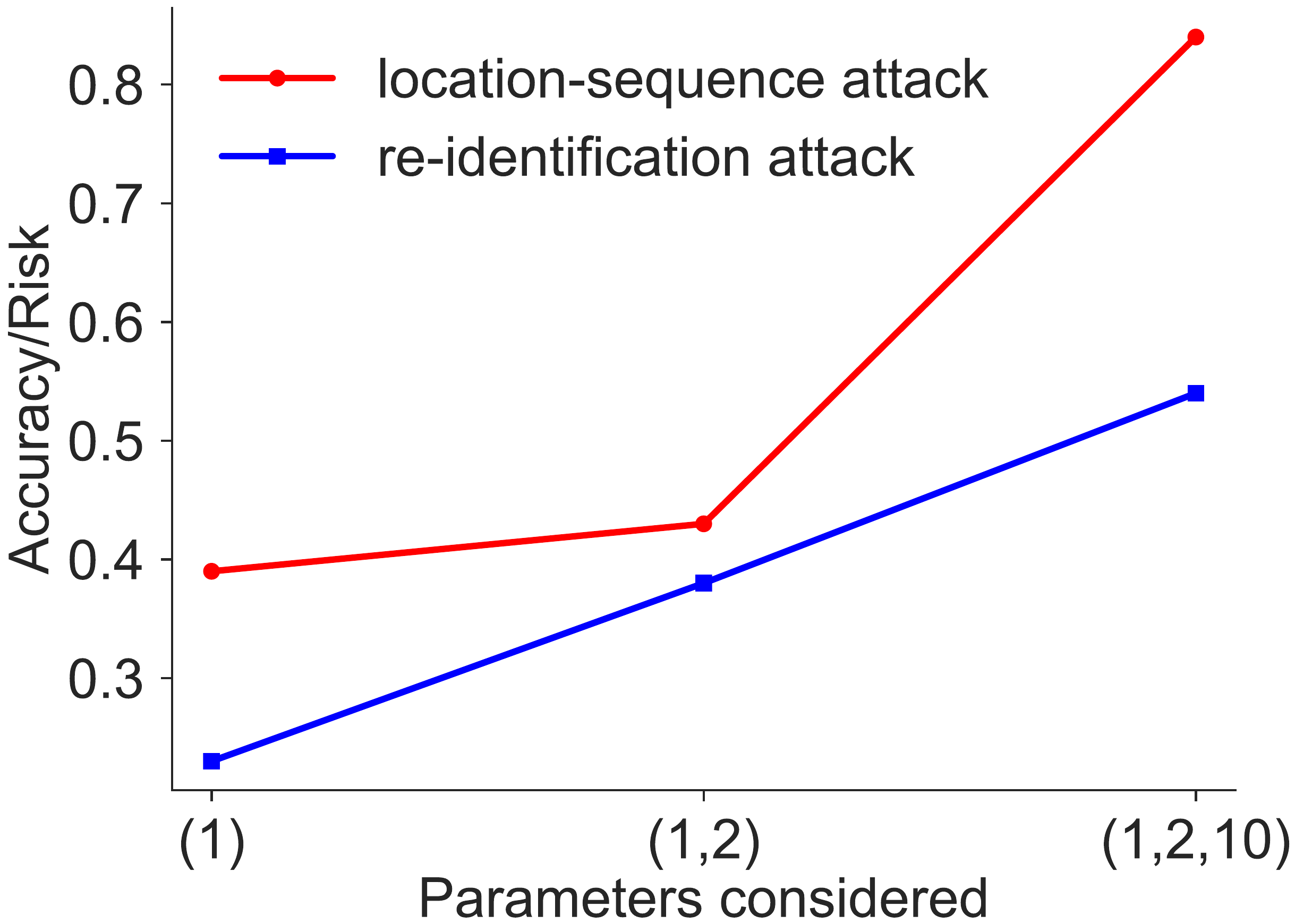}
        \caption{Adversarial attack accuracy}
        \label{fig:param_attack}
    \end{subfigure}
    \begin{subfigure}[b]{0.24\textwidth}
        \includegraphics[width=\textwidth]{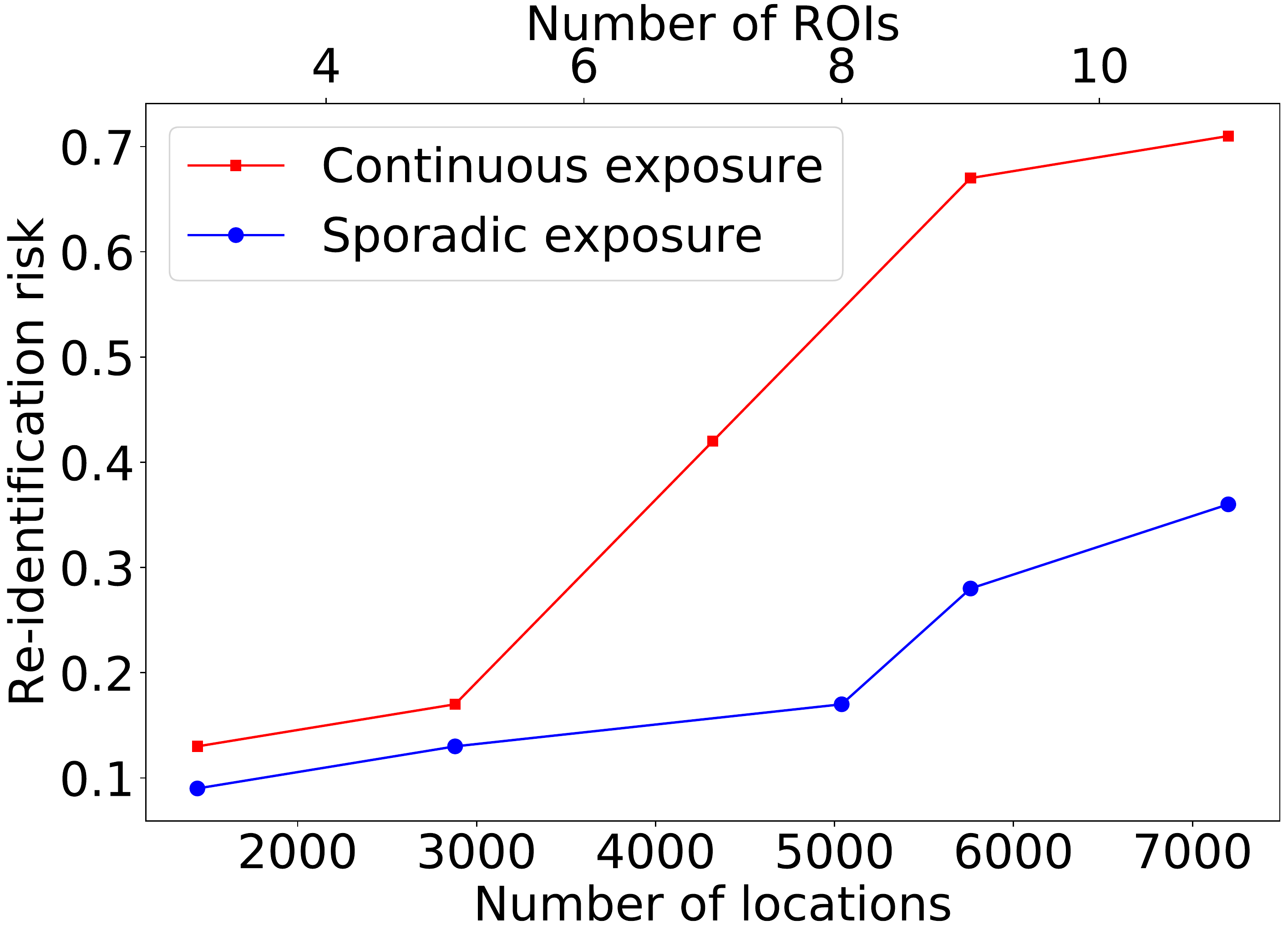}
        \caption{Adversarial attack risk}
        \label{fig:attack_con_spo}
    \end{subfigure}   
    \caption{Privacy analysis based on select users from Nokia dataset.}
    \label{fig:privacy_analysis}
\vspace{-15px}    
\end{figure}

The privacy by design approach cannot rely on measures such as differential privacy to perform privacy analysis, unlike the data concealing approaches.  
We follow the methodology specified by Shokri et al.~\cite{shokri2011quantifying} involving construction of a schedule consisting of an application, a LPPM (location privacy preserving mechanism), an attack and the evaluation metric. 
In our case, an application can be any continuous exposure LBS at the user's end and our LPPM is the minimization of the exposed locations via on-board processing. 
We consider two commonly used attacks: (i) location-sequence attack, and (2) re-identification attack.    
The success of these attacks depends on the adversary's prior knowledge, i.e. access to some traces of users or public information such as visited locations.   
Finally the user's privacy is quantified in terms of the correctness/incorrectness of the attacks by using the Location-privacy and mobility meter{\footnote{$LPM^2$: icapeople.epfl.ch/rshokri/lpm/doc/}} and privacy-lib.{\footnote{privacy-lib: github.com/pellungrobe/privacy-lib}}
In case of a privacy by design based system, we can clearly see (Figure~\ref{fig:attack_con_spo}) that by minimizing the locations shared with the third party services, we lower the adversaries prior knowledge, hence the risk of the attacks resulting in an increased user privacy as stated in~\cite{shokri2011quantifying}. 
Furthermore, by not relying on behavioral parameters, we lower the adversaries background knowledge contributing to enhanced user privacy as depicted in Figure~\ref{fig:param_attack}.
Here, we consider three parameters for evaluation: (1) $Max_{dist}$, (2) $Min_{time}$, and (3) $Min_{visit}$. 
We do not take service utility in to account as we assume a system based on a trusted computing environment, which does not compromise on service utility~\cite{Kulkarni2017PrivacyPreservingLS}. 
However, in case of techniques such as obfuscations or anonymization, pseudo-locations are used for the last hop, in which case the accuracy depends on the amount of distortion added to the user's true location.

\section{Related Work}
\label{sec:related_work}

\begin{figure}[!t]
\centering
\includegraphics[scale=0.45]{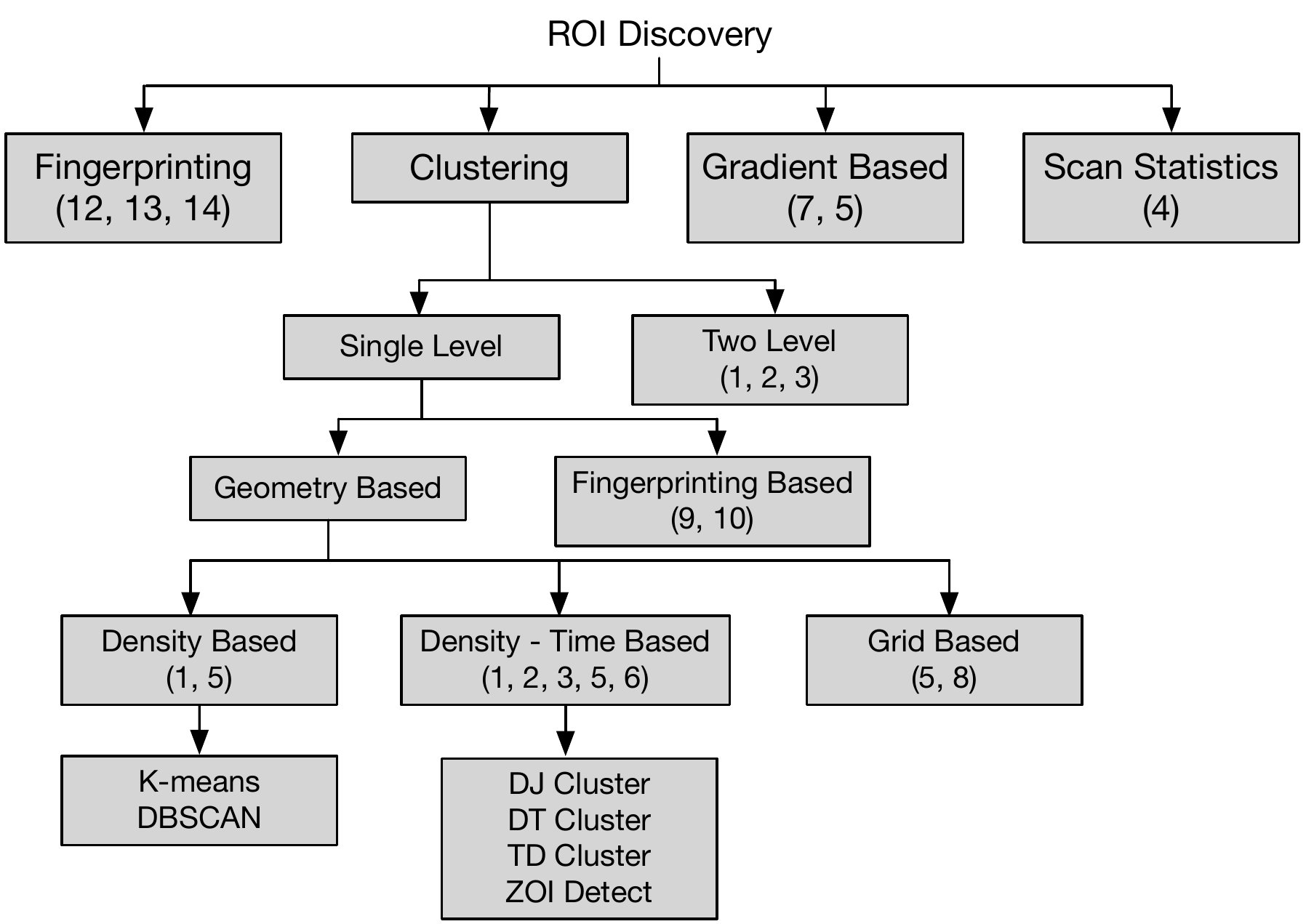}
\caption{Summary of ROI discovery techniques and their categorization. The numbers represent the dependent parameters from Table~\ref{tbl:ROI_parameters}.}
\label{fig:related_work}
\vspace{-15px}
\end{figure}

We present a new perspective on treating mobility trajectories as space-time signals to model human mobility in a computationally efficient manner. 
However, as a major aspect of our technique is ROI and transition discovery from the signals without relying on any behavioral parameters, we review the state of the art in this field. 
The existing techniques and their dependent parameter space is depicted in Figure~\ref{fig:related_work}. 

\vspace{3px}

{\noindent {\bf{Clustering.}}} An iterative approach for extracting the ROIs using clustering was proposed by Ashbrook et al.~\cite{2ashbrook2002learning}. 
The ROI granularity was improved by setting the spatiotemporal bounds derived by analyzing their variance with respect to the dependent values. 
Montoliu et al.~\cite{4montoliu2010discovering} proposed a two-level grid-based clustering approach, where the points are successively clustered in the temporal and spatial domain.    
Algorithms such as k-means~\cite{hartigan1979algorithm}, neighbor density-based clustering (DBSCAN)~\cite{citeulike:3509601} and time-density clustering~\cite{Gambs:2011:SMY:2019316.2019320} are used to detect clusters in spatiotemporal datasets, which are then considered as ROIs.  
Zheng et al.~\cite{7zheng2009mining} proposes a clustering-based ROI extraction technique where the parameters are estimated by observing the distribution of movement density. 

\noindent{\textbf{Fingerprinting}}. This technique relies on estimating a user's location fingerprint based on the detection of stable radio environments indicating a ROI.  
Farrahi et al.~\cite{3farrahi2011discovering} extracts ROIs by first forming a vector of the visible cell towers and then using repeatability rates and location transitions over time to filter out insignificant places.  
This technique is particularly applied to identify places, such as work and home.

\noindent{\textbf{Scan Statistics}}. This technique proposed by Fanaee-T et al.~\cite{6fanaee2014eigenvector} moves a cylinder of varying radii and height over a spatiotemporal space, where the surface covers the spatial dimension and height covers the temporal dimension. 
The cylinders are then sorted depending on the F-score, and an additional parameter called p-value is used as a threshold to filter insignificant places. 
The authors show the ability of this technique to handle complex data and reduce noise, at the cost of loosing the original shape of the ROI. 
However, spatial scan statistics is based on a frequentist view and does not depend on other priors like in Bayesian statistics.

\noindent{\textbf{Gradient Based}}. Thomason et al.~\cite{9thomason2016identifying} proposed a technique that combines the advantages of both k-means and DBSCAN.
It does not place a bound on visit duration or enforce any delay on trajectory points being considered, thus it can be operated in instantaneous time.
The thresholds used in this work are empirically derived and the results achieved have minimum bias due to the parameters.
Louail et al.~\cite{Louail2014FromMP} propose a technique to extract ROIs from trajectories belonging to a group of users without relying on the commonly used spatiotemporal bounds. 
However, they consider a group of points at a particular time, as a ROI if the density of users at that location is greater than a predefined threshold.  
\section{Conclusion}
\label{sec:future_work}
 
The paradigm shift towards cloud computing has encouraged LBS providers to deploy their infrastructure on untrusted cloud providers.   
This context has created several privacy and confidentiality issues by aggregating large amount of user location information in third-party datacenters.
Although, several techniques have been proposed to curb the location privacy leakage, a large gap still exists between the theoretical body of knowledge and the real-world applications.   
Furthermore, the new EU data protection regulations have imposed stringent restrictions on the volume of data aggregated, processed and stored at the service provider's end. 

In order to address these issues and facilitate the trend of on-board processing at the user's end, we have proposed a novel perspective on spatiotemporal computation by treating trajectories as space-time signals.  
We have leveraged the properties of these signals to reduce the computational complexity and power consumption.
We have presented {\textbf{Capstone}}, that illustrates this approach on mobility modeling task and shows that, not only do the signals preserve all the key knowledge contained in the trajectories but also formulate the mobility models with a high accuracy.
We have evaluated in depth the proposed technique by first analyzing it only from the signal processing perspective, and then verifying whether it satisfies already proven measures of human mobility.    
Our validation with the ground truth achieves precision and recall rates exceeding 80\% and achieves results on par with the conventional clustering approaches.   
We have performed the complexity and power consumption analysis by implementing {\textbf{Capstone}} on a DSP chip commonly present in many smartphones.
Furthermore, we have experimentally depicted the bias resulting from the stringent parameter bounds in the mobility modeling process and the associated privacy leakage. 
We have also demonstrated the suitability of our technique to extract ROIs from a larger variety of datasets and across different mobility behaviors.

\tiny{
\bibliographystyle{abbrv}
\bibliography{conference_main}
}

\end{document}